\documentclass[11pt]{article}
\usepackage{amsmath,bm,amsthm,bbm}
\usepackage{amssymb}
\usepackage{graphicx,psfrag,epsf}
\usepackage{enumerate}
\usepackage{natbib}
\usepackage{url}
\usepackage{float}
\usepackage{color}
\usepackage{placeins}
\usepackage{appendix}
\usepackage[normalem]{ulem}
\usepackage[dvipsnames]{xcolor}
\newtheorem{theorem}{Theorem}[section]
\newtheorem{definition}[theorem]{Definition}
\newtheorem{remark}[theorem]{Remark}
\newtheorem{proposition}[theorem]{Proposition}

\usepackage{fullpage}

\newcommand{\blind}{1}

\usepackage{xr-hyper}
\usepackage{hyperref}

\hypersetup{
  colorlinks=true,
  linkcolor=blue,
  citecolor=blue,
  filecolor=blue,
  urlcolor=blue,
}

\protect

\begin{document}

\def\spacingset#1{\renewcommand{\baselinestretch}%
{#1}\small\normalsize} \spacingset{1}

\spacingset{1} 


\if1\blind
{
  \title{Mixed Time Series Quasi-Likelihood Models for Uncovering Covid-19 Viral Load and Mortality Dynamics}
  \author{Kejin Wu\footnote{\texttt{e-mail}: \href{mailto:kwu8@luc.edu}{kwu8@luc.edu}}\\
    {\normalsize \it Department of Mathematics and Statistics, Loyola University Chicago, USA}\\
    Raanju R. Sundararajan\footnote{\texttt{e-mail}: \href{mailto:rsundararajan@mail.smu.edu}{rsundararajan@mail.smu.edu}}\\ 
    {\normalsize \it Department of Statistics and Data Science, Southern Methodist University, USA} \\
    Michel F.C. Haddad\footnote{\texttt{e-mail}: \href{mailto:m.haddad@qmul.ac.uk}{m.haddad@qmul.ac.uk}} \\
    {\normalsize \it Department of Business Analytics and Applied Economics, Queen Mary University of London, UK} \\
    Luiza S.C. Piancastelli\footnote{\texttt{e-mail}: \href{mailto:luiza.piancastelli@ucd.ie}{luiza.piancastelli@ucd.ie}} \\
    {\normalsize \it School of Mathematics and Statistics, University College Dublin, Republic of Ireland}\\
    Wagner Barreto-Souza\footnote{\texttt{e-mail}: \href{mailto:wagner.barreto-souza@ucd.ie}{wagner.barreto-souza@ucd.ie} (corresponding author)} \\
    {\normalsize \it School of Mathematics and Statistics, University College Dublin, Republic of Ireland}}
  \maketitle
} \fi

\if0\blind
{
  \bigskip
  \bigskip
  \bigskip
  \begin{center}
    {\LARGE Mixed Time Series Quasi-Likelihood Models for Uncovering Covid-19 Viral Load and Mortality Dynamics}
\end{center}
  \medskip
} \fi

\bigskip
\begin{abstract}
\noindent Accurate real-time monitoring of disease transmission is crucial for epidemic control, which has conventionally relied on reported cases or hospital admissions. Such metrics are frequently susceptible to delays in reporting, various forms of bias, and under-ascertainment. Cycle threshold values obtained from reverse transcription quantitative polymerase chain reaction offer a promising alternative, serving as a proxy for viral load. In this paper, we aim to jointly model the viral load and the number of deaths (mortality), which involves a continuous bounded and a count time series, and therefore, a proper mixed-type model is needed. This is the motivation to introduce a new mixed-valued time series quasi-likelihood (\texttt{MixTSQL}) model capable of analyzing multivariate time series of different types, like continuous, discrete, bounded, and continuous positive. The \texttt{MixTSQL} model only requires a mean-variance specification with no distributional assumptions needed, and allows for testing Granger causality. Statistical guarantees are provided to ensure consistency and asymptotic normality of the proposed quasi-maximum likelihood estimators. We analyze weekly viral load and Covid-19 death counts in São Paulo, Brazil, using our \texttt{MixTSQL} model, which not only establishes the temporal order in which viral load Granger-causes mortality but also offers a comprehensive joint statistical analysis.

\end{abstract}

\noindent%


\section{Introduction}
\label{sec:intro}

Events of epidemics and pandemics represent widespread occurrences of infectious diseases that increase death rates across regions. The frequency of such events has grown over the past hundred years, driven by factors such as intensified global mobility, urban expansion, and increased exploitation of natural ecosystems. In response, considerable policy efforts have emphasized the importance of detecting and containing disease outbreaks with pandemic potential, alongside investments to improve public health preparedness and system resilience \citep{madhav2017pandemics}. The global Covid-19 pandemic was precipitated by the Severe Acute Respiratory Syndrome Coronavirus 2 (SARS-CoV-2). The virus propagates between human hosts via respiratory aerosol particles. Considering the number of related death counts, Brazil is the second most affected country in the world. During the peak of the Covid-19 crisis in Brazil, daily mortality figures exceeded $4,000$ fatalities attributable to the disease. Even after the most critical moments of the global pandemic, many streams of virus have emerged in different parts of the world. Recent examples include the variant Nimbus (NB.1.8.1), and a new subvariant of the Omicron known as Stratus (XFG), among others \citep{geddes2025nimbusstratus}.

 The ability to generate reliable epidemiological predictions and causal relationships is highly desirable. The timely implementation of evidence-based policy interventions (e.g., social distancing protocols) commonly mitigates mortality rates. There exists a substantial theoretical justification to posit that heterogeneity in individual transmission capability may profoundly impact the progression and dynamics of a pandemic episode \citep{lloyd2005superspreading}. Conventionally, epidemiological surveillance relies upon reported cases, hospital admissions, or related variables (e.g., positivity rate, hospital average stay). However, such metrics are frequently susceptible to relevant caveats and limitations. We assess the utility of viral load estimates derived from reverse transcription quantitative polymerase chain reaction (RT-qPCR) cycle threshold (Ct) values as a promising alternative for monitoring epidemic dynamics \citep{lim2024nowcasting}. It also merits acknowledgment that viral load constitutes a universally applicable measure of pathogenic progression. Therefore, the present work is not exclusively pertinent to SARS-CoV-2 infections, being equally useful considering future episodes of epidemics or pandemics caused by a diverse range of viruses.

In this paper, we aim to study and understand the joint dynamics of the Covid-19 viral load, which is defined in terms of Ct, and its associated mortality (number of deaths). We consider weekly data from São Paulo, Brazil, with the sampling period spanning from the 20th March 2020 to the 22nd May 2022. This data involves a continuous bounded (viral load) and a count (mortality) time series, and therefore, a proper mixed-type model is needed. This is the motivation to introduce a new mixed-valued time series quasi-likelihood model (in short \texttt{MixTSQL}) capable of analyzing multivariate time series of different types, like continuous, discrete, bounded, and continuous positive-valued, and which only requires a mean-variance specification instead of a full distributional assumption. 
 
The study of interactions among time series has been of interest in areas such as finance, economics, and neuroscience since the seminal paper by \cite{gra1969}, where the concept of Granger causality was introduced. The majority of the approaches for testing Granger causality relies on the linear vector autoregressive (VAR) model, which relies on the multivariate normal distribution. On the other hand, time series data may be of a different nature such as counts, positive continuous, or bounded valued. The causality problem involving just count time series has been addressed by \cite{cheandlee2017}, \cite{leeandlee2019}, and \cite{piaetal2023}, while \cite{tanetal2021} proposed a test for dealing with categorical time series. For a review that accounts for the most important developments and recent advances in Granger causality, we refer to the work by \cite{shofox2022}. One of the few papers handling mixed-valued time series is due to \cite{piaetal2024}, where a class of mixed-valued time series generalized linear models (GLMs) was proposed. One of the requirements of these models is that the conditional distributions (given past observations) must belong to the exponential family, and therefore cannot be used to analyze our data since one of the components (viral load) is bounded. In fact, our formulation can be seen as a generalization of the models proposed by \cite{piaetal2024} in the same sense quasi-likelihood models extend GLMs.

We highlight that an important and rather distinctive advantage of the models introduced in the present work is that only a mean-variance specification needs to be specified. No specific distributional assumptions are needed, which implies more flexibility when compared to fully parametric models. Under the assumption that the mean function's structural specification is correct, we provide theoretically consistent estimation for the model coefficients. Another highly relevant feature is that we are able to test Granger causality between two time series that can be binary, counts, continuous, positive, and bounded-valued.

We summarize the main contributions of this paper in what follows.

\begin{itemize}
    
\item The present work focuses on the use of intrinsic information contained within the cycle threshold values (Ct) derived from the RT-qPCR assays, through viral load, which is defined as one minus the standardized Ct, and the study of its relationship with mortality counts. With this challenge in mind, a new quasi-likelihood model, termed as \texttt{MixTSQL}, is introduced and designed for analyzing the mixed-valued time series Covid-19 data. Instead of assuming a fully parametric multivariate time series model, \texttt{MixTSQL} models only impose assumptions on the first two cumulants of the time series. Then, a quasi-likelihood approach is developed for estimation and inference purposes. As long as the structural assumption on the mean function is correct, consistency in mean estimation is obtained. Statistical guarantees to ensure consistency and asymptotic normality of the proposed quasi-likelihood estimators are provided.

\item Very few works consider testing Granger causality in mean and variance simultaneously. In the present work, we provide the necessary and sufficient conditions to check the Granger causality in mean and variance simultaneously. This test is critically important while understanding the lead-lag effect of viral load on subsequent mortalities. The individual definitions of Granger causality with respect to the mean or with respect to the variance can be found in \citep{gra1969} and \cite{graetal1986}, respectively; see also \cite{stock1989interpreting, lee1992causal, maansson2009granger, ccevik2018oil} for a list of wide-ranging applications.

\item Extensive simulation experiments are conducted to study the performance of the new quasi-likelihood estimation technique. The simulation settings are carefully designed to closely mimic the Covid-19 data variables under consideration. It is well known that standard error computations based on theory-driven analytical expressions need not be accurate in finite sample situations; see \cite{barreto2025time} for a recent work comparing theoretical and bootstrap approaches in the context of count time series models. Our simulation studies, however, reveal that both theoretical and bootstrap approaches work well in finite sample settings. We also discuss connections between the asymptotic results on the estimators and the empirical results from this simulation study.

\item We investigate the impact of viral load on the modeling and forecasting of mortality, while also testing for Granger causality. This novel approach offers valuable insights for future outbreaks of viral epidemics or pandemics. We perform an out-of-sample forecasting exercise, which shows that the \texttt{MixTSQL} model yields a lower root mean forecasting error compared to a standard Gaussian linear model. Moreover, our model provides strong evidence of Granger causality from viral load to mortality counts.
\end{itemize}

The paper unfolds as follows. In Section \ref{sec:data}, we describe the Covid-19 data from Brazil in detail. In Section \ref{sec:model}, we introduce our class of \texttt{MixTSQL} models for analyzing mixed-valued time series data. This section also includes model inference based on the quasi-likelihood approach, statistical guarantees to ensure consistency and asymptotic normality of the proposed estimators, and a Granger causality test. Simulation experiments and performance assessment results are reported in Section \ref{sec:simulation}. A full statistical data analysis of the Covid-19 viral load and mortality is provided in Section \ref{sec:application}. Section \ref{sec:conclusions} contains our concluding remarks, including possible future research. This paper contains Supplementary Material with all technical proofs of theorems and propositions.

\section{Covid-19 data description}
\label{sec:data}

Although daily reported positive cases constitute the primary surveillance measure for Covid-19 incidence, these data are acknowledged to exhibit relevant caveats and limitations - e.g., reporting delays, selection bias, under-ascertainment of actual infection rates \citep{flaxman2020estimating}. Moreover, its quantification of infectivity is restricted to a dichotomous assessment of an individual's contagiousness level. This binary characterization constitutes a significant methodological constraint in conducting meaningful and timely prognostic analyses.

Ct values consist of a continuous variable that is a proxy of viral load, indicating the amount of viral genetic material that is present in a patient sample. Despite its potential adoption for monitoring and mitigation purposes in cases of epidemics and pandemics, Ct values are mostly not publicly accessible. A low Ct value reflects a high concentration of viral genetic material, which is typically associated with a higher risk of infectivity. Conversely, a high Ct value reflects a low concentration of viral genetic material, which is typically associated with a lower risk of infectivity \citep{england2020understanding}. Notwithstanding inter-individual variability, specimen heterogeneity, and platform-dependent differences in assay performance, Ct values offer a probabilistic indicator of the time elapsed since infection acquisition. Accumulating evidence suggests a temporal association between epidemic trajectory and the mean Ct values observed in SARS-CoV-2–positive specimens \citep{kutta2025detection, hay2021estimating}.

An expanding epidemic or pandemic, characterized by a predominance of recent infections, exhibits higher viral loads and lower mean Ct values at the population level. Conversely, a declining epidemic or pandemic, in which a greater proportion of infections are in later stages, demonstrates lower viral loads and correspondingly elevated mean Ct values. Moreover, previous studies have reported a positive association between viral load and patients at higher risk of severe outcomes, including death - e.g., \cite{magleby2021impact}, \cite{faico2020higher}. Drawing on this insight, we introduce the methodological framework detailed in the subsequent section. It is worth mentioning that works involving viral load data are still scarce due to the challenges in collecting such datasets. Thus, more studies are needed to confirm this positive association - preferably, exploring larger sample sizes such as the one used in the present study.

The dataset explored in this work comprises raw PCR test results, including Ct values for SARS-CoV-2, obtained from a leading clinical diagnostic laboratory in Brazil (DB Diagnósticos). This dataset is complemented with a public related variable, namely the daily count of deaths attributed to the Covid-19. The dataset contains 342,699 individual PCR testing records. The sampling period spans from the 20th March 2020 to the 22nd May 2022. While specimens were collected across around 2,100 testing sites nationwide, all analyses were conducted at a centralized laboratory facility located in São Paulo. The Ct time series is bounded and we standardized it to the unit interval $(0,1)$. Then, the viral load time series is defined as $\mbox{viral load}=1-\mbox{standardized}(C_t)$. 

\begin{figure}[ht!]
    \centering
    \includegraphics[width=0.8\linewidth]{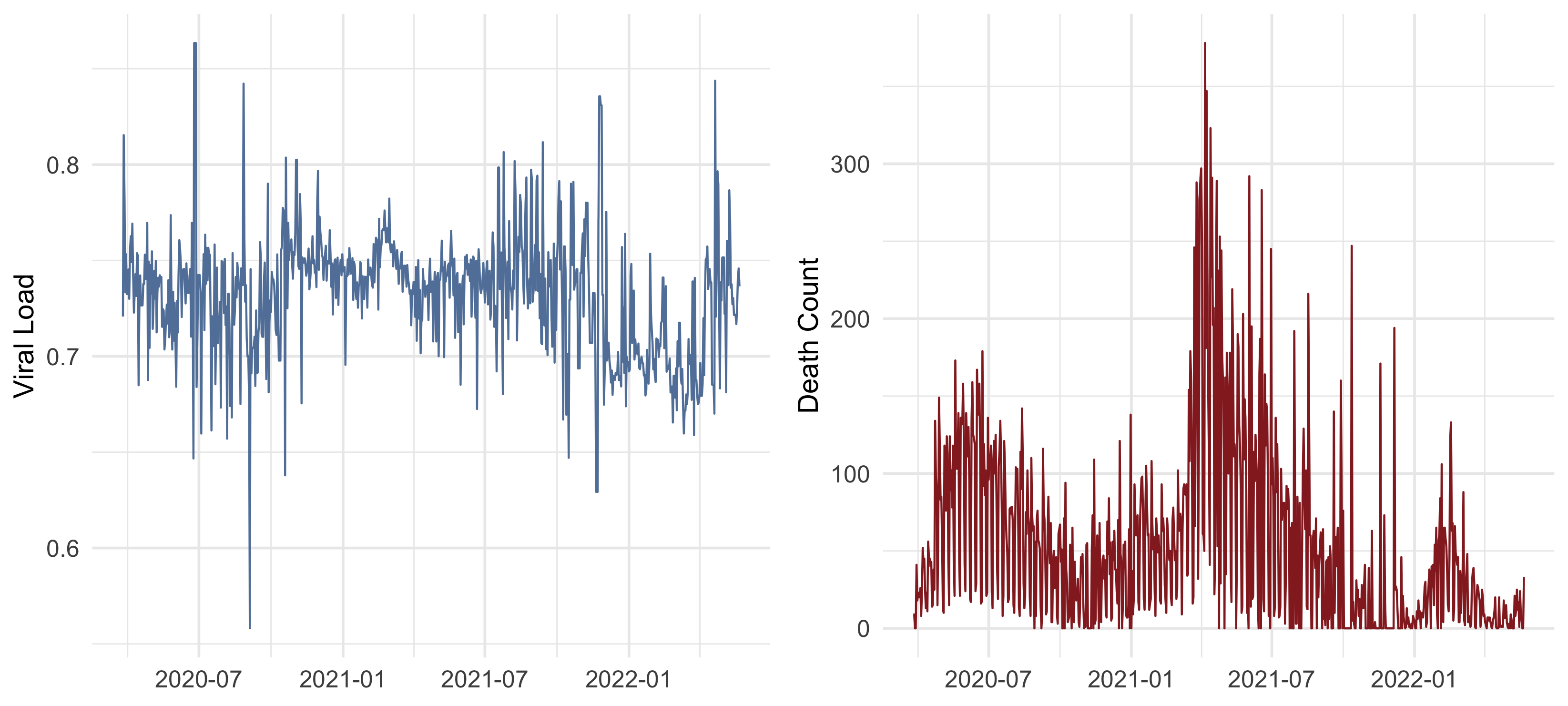}
    \caption{Daily observations of viral load and death counts in São Paulo, Brazil, from 2020-03-26 to 2022-05-22.}
    \label{fig:daily}
\end{figure}

The viral load and mortality bivariate mixed-valued time series, observed at the daily frequency in São Paulo, Brazil, is presented in Figure \ref{fig:daily}. The data at the daily frequency show a number of zero or close to zero deaths and high volatility, which brings difficulties to modeling. Not only does this behavior cause numerical instability, but also does not reflect reality. Zero daily deaths are most often due to a delay in reporting, occurring especially on weekends.

\begin{figure}[ht!]
    \centering
    \includegraphics[width=0.8\linewidth]{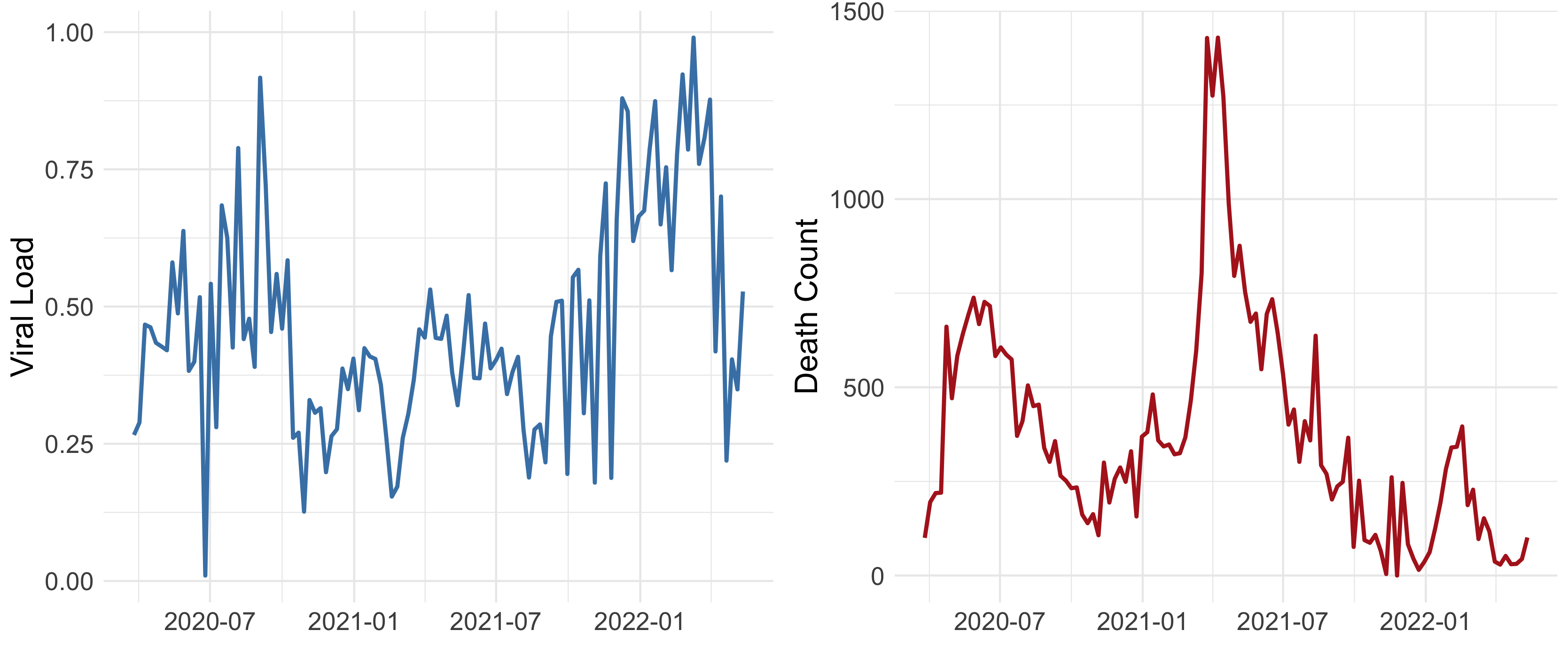}
    \caption{Weekly observations of viral load and death counts in São Paulo, Brazil, from 2020-03-26 to 2022-05-22.}
    \label{fig:weeklyseries}
\end{figure}

Therefore, we aggregate and analyze the two time series at a weekly frequency. Compared to the daily frequency data, the weekly data is preferable considering that commonly there are more PCR tests performed at the beginning of each week as a consequence of the fact that people get together more often   during weekends. As a result of weekly aggregation, the number of zero deaths decreases significantly and, consequently, the associated volatility, as illustrated in Figure \ref{fig:weeklyseries}.

In the present work, the analyses are performed at the weekly frequency of the data. Each transformed trajectory comprises of $n=112$ observations. It must also be noted that almost all existing studies exploring pandemic dynamics using Ct values work with a substantially smaller sample size compared to the present work; see \cite{hay2021estimating}, \cite{yagci2020relationship} for examples.

The primary objective of this paper is to jointly model viral load and mortality in order to better understand their dynamic relationship and investigate properties such as Granger causality. Given the mixed-type nature of the time series data and the lack of proper models existing in the literature to handle this problem, a novel methodological approach is required, which is introduced in the following section.

\section{Mixed Time Series Quasi-Likelihood Models}\label{sec:model}

In this section, we present a novel methodology for analyzing Covid-19 viral load dynamics and its association with related mortality. Our approach is grounded in a quasi-likelihood framework and extends the mixed time series generalized linear models introduced by \cite{piaetal2024}, which are based on exponential family formulations. This extension is essential for our application, as it accommodates key features of the Covid-19 time series data considered in this study, particularly the joint modeling of bounded and count-valued bivariate time series.

\subsection{Model specification}

We begin by introducing an important ingredient of our models. We consider the class of quasi-likelihood (QL) models proposed by \cite{wed1974}, which is defined by specifying the first two cumulants of a random variable (say $Y$), the mean $\mu$ and the variance $\mbox{Var}(Y)=\phi V(\mu)$, plus a log-quasi-likelihood function assuming the form
\begin{eqnarray}\label{qlfunction}
Q(y;\mu)=\dfrac{1}{\phi}\int_y^\mu\dfrac{y-\omega}{V(\omega)}d\omega,	
\end{eqnarray}	 
where $\phi>0$ is a dispersion parameter. The following notation is considered in this case, which implicitly depends on the variance function: $Y\sim\mbox{QL}(\mu,\phi)$.
A notable advantage of the QL modeling approach is that there are more possibilities to specify the variance function than in the exponential family. This allows, for example, to handle continuous bounded-valued data. The QL models contain the exponential family (GLM when considering covariates) as a special case, depending on the variance function specification.

Let $\{Y_{1\,t}\}_{t\in\mathbb N}$ and $\{Y_{2\,t}\}_{t\in\mathbb N}$ be two time series and denote the sigma-algebras $\mathcal F^{(1)}_{t-1}=\sigma(Y_{1\,t-1},Y_{1\,t-2},\ldots)$, $\mathcal F^{(2)}_{t-1}=\sigma(Y_{2\,t-1},Y_{2\,t-2},\ldots)$ and $\mathcal F^{(1,2)}_{t-1}=\sigma(\mathcal F^{(1)}_{t-1},\mathcal F^{(2)}_{t-1})$. We now define our class of bivariate mixed time series models as follows.

\begin{definition}\label{ts_granger_model}[\texttt{MixTSQL} model]
	The mixed-valued time series quasi-likelihood model (\texttt{MixTSQL}) is defined by the time series vector $\{(Y_{1\,t},Y_{2\,t})\}_{t\in\mathbb N}$, with $Y_{1\,t}$ and $Y_{2\,t}$ conditionally independent given $\mathcal F^{(1,2)}_{t-1}$ for all $t$, and satisfying $Y_{1\,t}|\mathcal F^{(1,2)}_{t-1}\sim\mbox{QL}^{(1)}(\mu_{1\,t},\phi_1)$ and $Y_{2\,t}|\mathcal F^{(1,2)}_{t-1}\sim\mbox{QL}^{(2)}(\mu_{2\,t},\phi_2)$, with 
	\begin{eqnarray}
		\nu_{2\,t}&\equiv& g_2(\mu_{2\,t})\equiv \beta_0^{(2)}+\sum_{i=1}^p \beta_i^{(2)}\widetilde Y_{2\, t-i}+\sum_{l=1}^k \gamma^{(2)}_l \widetilde Y_{1\, t-l},\label{eq2_mu2}\\
				\nu_{1\,t}&\equiv& g_1(\mu_{1\,t})\equiv \beta_0^{(1)}+\sum_{i=1}^r \beta_i^{(1)}\widetilde Y_{1\, t-i}+\sum_{l=1}^s \gamma^{(1)}_l \widetilde Y_{2\, t-l},\label{eq1_mu1}
	\end{eqnarray}	
where $g_1(\cdot)$ and $g_2(\cdot)$ are link functions assumed to be continuous, invertible, and twice differentiable, with $\widetilde Y_{1\, t}=T_1(Y_{1\, t})$ and $\widetilde Y_{2\, t}=T_2(Y_{2\, t})$ being adequate transformations of the original time series, $\beta_0^{(1)},\ldots,\beta_r^{(1)}$, $\beta_0^{(2)},\ldots,\beta_p^{(2)}$, $\gamma_1^{(1)},\ldots,\gamma_s^{(1)}$, and $\gamma_1^{(2)},\ldots,\gamma_k^{(2)}$ are real-valued parameters, and $\phi_1$ and $\phi_2$ are dispersion parameters.
\end{definition}

\begin{remark}
	The transformed time series $\widetilde Y_{1\, t}$  and $\widetilde Y_{2\, t}$ in (\ref{eq2_mu2}) and (\ref{eq1_mu1}), respectively, via $T_1(\cdot)$ and $T_2(\cdot)$ are necessary since we are modeling the transformed mean-related parameters. In general, we will consider $T_1(y)=g_1(y)$ and $T_2(y)=g_2(y)$ unless some slight modification is necessary such as in the count time series case. More specifically, if $Y_{2\, t}$ is a time series of counts and $g_2(y)=\log y$, we take $T_2(y)=\log(y+1)$ because the log-function is not be well-defined at $y=0$. This transformation technique is inspired by the log-linear INGARCH models by \cite{foktjo2011}. The choices for the transformed time series being the link functions or slight modifications of them will keep $\nu_{1\,t}$ and $\nu_{2\,t}$ in the same scales of $\widetilde Y_{1\, t}$ and $\widetilde Y_{2\, t}$, respectively. The superscripts (1) and (2) in $\mbox{QL}^{(1)}$ and $\mbox{QL}^{(2)}$ imply that different QL models can be used for $Y_{1\,t}$ and $Y_{2\,t}$.
\end{remark}

One of our aims is to test if $\{Y_{1\,t}\}_{t\in\mathbb N}$ causes to $\{Y_{2\,t}\}_{t\in\mathbb N}$ in mean \citep{gra1969}, that is
\begin{eqnarray}\label{granger_in_mean}
\mbox{Pr}\left(E(Y_{2\,t}|\mathcal F^{(2)}_{t-1})\neq E(Y_{2\,t}|\mathcal F^{(1,2)}_{t-1})\right)>0.
\end{eqnarray}

\noindent Throughout this paper, we consider the Granger causality test for whether $\{Y_{1,t}\}_{t \in \mathbb{N}}$ Granger-causes $\{Y_{2,t}\}_{t \in \mathbb{N}}$. The test for the reverse direction proceeds analogously.

The choice for the variance function will dictate how the variance will depend on the mean. As a further consequence, the series $\{Y_{1\,t}\}_{t\in\mathbb N}$ can have a causal effect on the series $\{Y_{2\,t}\}_{t\in\mathbb N}$ in both the  mean and variance simultaneously. According \cite{graetal1986}, $\{Y_{1\,t}\}_{t\in\mathbb N}$ causes to $\{Y_{2\,t}\}_{t\in\mathbb N}$ in variance if
\begin{eqnarray}\label{granger_in_variance}
\mbox{Pr}\left(\mbox{Var}(Y_{2\,t}|\mathcal F^{(2)}_{t-1})\neq \mbox{Var}(Y_{2\,t}|\mathcal F^{(1,2)}_{t-1})\right)>0.
\end{eqnarray}
More specifically, the inequality in (\ref{granger_in_variance}) is
\begin{eqnarray}\label{granger_in_variance2}
\mbox{Pr}\left(E\left[\left(Y_{2\,t} - E(Y_{2\,t}|\mathcal F^{(1,2)}_{t-1})\right)^2 \Bigg|\mathcal F^{(2)}_{t-1}\right]\neq E\left[\left(Y_{2\,t} - E(Y_{2\,t}|\mathcal F^{(1,2)}_{t-1})\right)^2 \Bigg|\mathcal F^{(1,2)}_{t-1}\right]\right)>0.
\end{eqnarray}
Readers can refer to \cite{guoetal2014} for simultaneous causality testing in the factor double autoregressions problem. Following their approach, we consider the following.
\begin{eqnarray}\label{granger_in_simultaneous}
\mbox{Pr}\left(E\left[\left(Y_{2\,t} - E(Y_{2\,t}|\mathcal F^{(2)}_{t-1})\right)^2 \Bigg|\mathcal F^{(2)}_{t-1}\right]\neq E\left[\left(Y_{2\,t} - E(Y_{2\,t}|\mathcal F^{(1,2)}_{t-1})\right)^2 \Bigg|\mathcal F^{(1,2)}_{t-1}\right]\right)>0.
\end{eqnarray}

We now provide a proposition that shows that verifying inequality (\ref{granger_in_simultaneous}) is sufficient to test for Granger causality in mean and variance simultaneously. 
\begin{proposition}\label{Prop:inequality} Inequality (\ref{granger_in_simultaneous}) holds if at least one of the inequalities in (\ref{granger_in_mean}) and (\ref{granger_in_variance}) holds.
\end{proposition}

In practice, it is common to assume a data-generating model to capture the underlying dynamics of observed time series, such as ARMA models for univariate data or VAR models for multivariate settings. Subsequently, the Granger causality test can be performed by inferring on the values of certain model coefficients. However, these classical models often impose too many restrictions such as  normality and continuous-valued time series, besides being susceptible to model misspecification.  
The proposed class of \texttt{MixTSQL} models is robust to model misspecification due to its quasi-likelihood approach and offers a flexible framework for testing Granger causality among time series of different types, including continuous, positive, count, bounded, and binary variables.

Under the proposed \texttt{MixTSQL} model defined in Definition \ref{ts_granger_model}, testing for Granger causality of the time series ${Y_{1,t}}$ on ${Y_{2,t}}$ in both mean and variance is equivalent to testing whether the Granger causality parameters $\gamma_{l}^{(2)}$ are nonzero for 
$l=1,\ldots,k$, as established in Proposition \ref{Prop:GrangerTest}. We now introduce a technical assumption required to prove such a result. A Granger causality test will be discussed in Subsection \ref{sec:causality_test}.

\begin{itemize}
    \item A1. The $\sigma$-algebras $\sigma(\sigma(Y_{2\, t}), \mathcal F^{(2)}_{t-1})$ and $\mathcal F^{(1)}_{t-1}$ are independent when $\gamma_{l}^{(2)}=0 \;\; \forall \; l \in \{1,2,\hdots,k\}$. 
\end{itemize}

\begin{remark}
    Assumption A1 implies that the random variable $Y_{2\, t}$ and the sequence $\{ Y_{1\, t-1}, $ $Y_{1\, t-2}, \ldots\}$ are conditionally independent given $\mathcal F^{(2)}_{t-1}$ when $\gamma_{l}^{(2)}=0 \;\; \forall \; l \in \{1,2,\hdots,k\}$. 
\end{remark}

\begin{proposition}\label{Prop:GrangerTest} Under Assumption A1, the inequality in (\ref{granger_in_simultaneous}) holds if and only if  $\gamma_{l}^{(2)} \neq 0$ for some $l \in \{1,2,\hdots,k\}$. 
\end{proposition}

Assume that $\{(y_{1\, t},y_{2\, t})\}_{t=1}^n$ is a realization of a mixed-valued time series from the \texttt{MixTSQL} model described in Definition \ref{ts_granger_model}. Define the parameter vector $\boldsymbol\theta=(\bm \theta^{(1)},\boldsymbol\theta^{(2)})^\top$. Note that $\phi_1$ and $\phi_2$ are the nuisance parameters. The quasi log-likelihood function, say $\widetilde Q(\boldsymbol\theta)$, is given by
\begin{eqnarray} \label{eq:Q-function-split}
\widetilde Q(\boldsymbol\theta)=\sum_{t=m+1}^n\left\{Q_1(y_{1\,t};\mu_{1\,t})+Q_2(y_{2\,t};\mu_{2\,t})\right\},
\end{eqnarray} 
where $m=\max(p,q,r,s)$, $Q_j(\cdot;\cdot)$ assumes the form from (\ref{qlfunction}) with series specific function $V(\cdot) :=V_j(\cdot)$, for $j=1,2$.

The quasi-maximum likelihood estimator (QMLE) of $\boldsymbol\theta$ (conditional on $\{(y_{1\,i},y_{2\,i})\}_{i=1}^m$) is given by $\widehat{\boldsymbol\theta}:=\arg\min_{\boldsymbol\theta}-\widetilde Q(\boldsymbol\theta)$. The quasi score function associated with $\widetilde Q(\boldsymbol\theta)$ is
${\bf U}(\boldsymbol\theta)\equiv\dfrac{\partial \widetilde{Q}(\boldsymbol\theta)}{\partial\boldsymbol\theta}=\displaystyle\sum_{t=m+1}^n {\bf U}_t(\boldsymbol\theta)$, where 
\begin{eqnarray}\label{eq:score}
{\bf U}_t(\boldsymbol\theta)&=&\left(\dfrac{1}{\phi_1}\dfrac{y_{1t}-\mu_{1t}}{V(\mu_{1t})}\dfrac{1}{g_1'(\mu_{1t})}\dfrac{\partial\nu_{1t}}{\partial\bm \theta^{(1)}},\dfrac{1}{\phi_2}\dfrac{y_{2t}-\mu_{2t}}{V(\mu_{2t})}\dfrac{1}{g_2'(\mu_{2t})}\dfrac{\partial\nu_{2t}}{\partial\boldsymbol\theta^{(2)}}\right)^\top,\\
\dfrac{\partial\nu_{1t}}{\partial\bm \theta^{(1)}}&=&(1,\widetilde{Y}_{1\, t-1},\ldots,\widetilde{Y}_{1\, t-r},\widetilde{Y}_{2\, t-1},\ldots,\widetilde{Y}_{2\, t-s})^\top,\nonumber\\
\dfrac{\partial\nu_{2t}}{\partial\boldsymbol\theta^{(2)}}&=&(1,\widetilde{Y}_{2\, t-1},\ldots,\widetilde{Y}_{2\, t-p},\widetilde{Y}_{1\, t-1},\ldots,\widetilde{Y}_{1\, t-k})^\top,
\end{eqnarray}
with $\bm \theta^{(1)}=(\beta_0^{(1)},\beta_1^{(1)},\ldots,\beta_r^{(1)}, \gamma_1^{(1)},\ldots,\gamma_s^{(1)})^\top$ and $\boldsymbol\theta^{(2)}=(\beta_0^{(2)},\beta_1^{(2)},\ldots,\beta_p^{(2)}, \gamma_1^{(2)},\ldots,\gamma_k^{(2)})^\top$.

If the dispersion parameters are unknown, they can be estimated via method of moments. Specifically, these estimators assume the forms
\begin{eqnarray}\label{eq:phi_est}
\widehat\phi_1=\dfrac{\displaystyle\sum_{t=m+1}^n\left(Y_{1\,t}-\widehat\mu_{1\,t}\right)^2}{\displaystyle\sum_{t=m+1}^nV_1\left(\widehat\mu_{1\,t}\right)}	
\quad\mbox{and}\quad
\widehat\phi_2=\dfrac{\displaystyle\sum_{t=m+1}^n\left(Y_{2\,t}-\widehat\mu_{2\,t}\right)^2}{\displaystyle\sum_{t=m+1}^nV_2\left(\widehat\mu_{2\,t}\right)},	
\end{eqnarray}	
where $\widehat\mu_{1\,t}$ and $\widehat\mu_{2\,t}$ are the quasi-likelihood estimators of $\mu_{1\,t}$ and $\mu_{2\,t}$, respectively. In the following subsection, we provide statistical guarantees to ensure consistency and asymptotic normality of the quasi-likelihood estimators.

\subsection{Statistical guarantees and asymptotics}\label{subsec:asymptotics}

We first consider the consistency of the quasi-maximum likelihood estimator (QMLE) $\widehat{\boldsymbol\theta}$. To simplify the notation, we denote $Q_1(Y_{1\,t};\mu_{1\,t})$ and $Q_2(Y_{2\,t};\mu_{2\,t})$ from \eqref{eq:Q-function-split} by $Q_{1\, t}(\bm \theta^{(1)})$ and $Q_{2\, t}(\bm \theta^{(2)})$, respectively, where the parameter vector $\bm \theta = (\boldsymbol\theta^{(1)}, \bm \theta^{(2)})^\top$, with $\bm \theta^{(1)}$ and $\boldsymbol\theta^{(2)}$ as defined above. This enables writing $\widetilde Q(\boldsymbol\theta)= \sum_{t=m+1}^n\{Q_{1\, t}(\bm \theta^{(1)}) + Q_{2\, t}(\bm \theta^{(2)})\}$. Note that, from the \texttt{MixTSQL} model's Definition \ref{ts_granger_model}, $\nu_{1\,t}$ and $\nu_{2\,t}$ depend exclusively on $\bm \theta^{(1)}$ and $\bm \theta^{(2)}$, respectively. This allows us to write these two quantities as $\nu_{1\,t}(\bm \theta^{(1)})$ and $\nu_{2\,t}(\bm \theta^{(2)})$. Next, a few technical assumptions are provided and these are needed to establish consistency of the proposed estimators. 
\begin{itemize}
    \item A2. The series $ \{\mathbf{Y}_t =  (Y_{1\,t},Y_{2\,t})\}$ is stationary and ergodic.
    \item B1. The true value $\bm \theta_0 := (\bm \theta^{(1)}_{0}, \bm \theta^{(2)}_{0})^\top$ is an interior point in $\Theta := \Theta_1 \times\Theta_2$ which is a compact set in $\mathbb R^{r+s+1}\times\mathbb R^{p+k+1}$. 
    \item B2. $\nu_{1\,t}(\bm \theta^{(1)}_{0}) = \nu_{1\,t}(\bm \theta^{(1)})$ and $\nu_{2\,t}(\bm \theta^{(2)}_{0}) = \nu_{2\,t}(\bm \theta^{(2)})$ if and only if $\bm \theta^{(1)}_{0} = \bm \theta^{(1)}$ and $\bm \theta^{(2)}_{0} = \bm \theta^{(2)}$, respectively.
    \item C1. $E\left[\sup _{\bm \theta \in \Theta}\left|Q_{1\, t}(\bm \theta^{(1)}) + Q_{2\, t}(\bm \theta^{(2)})\right|\right]<\infty$.
    \item C2. $E\left[-Q_{1\, t}(\bm \theta^{(1)}) - Q_{2\, t}(\bm \theta^{(2)})\right]$ has a unique minimum at $\bm \theta_0$.
\end{itemize}

\begin{remark}
    It is important to note that Theorem \ref{Theorem:consistency} could still hold even without the stationarity condition from Assumption A2. For example, we can assume $\frac{1}{n-m} \sum_{t=m+1}^n [\inf _{\boldsymbol{\theta} \in B_{\eta_0}\left(\boldsymbol{\theta}_*\right)}-Q_t(\boldsymbol{\theta}) - $ $ E \left(\inf _{\boldsymbol{\theta} \in B_{\eta_0}\left(\boldsymbol{\theta}_*\right)}-Q_t(\boldsymbol{\theta})\right)] \to 0$ for $\forall~\bm{\theta}_*$, which is similar to Kolmogorov’s strong law of large numbers, where $B_\eta\left(\boldsymbol{\theta}_*\right)=\left\{\boldsymbol{\theta} \in \Theta:\left\|\boldsymbol{\theta}-\boldsymbol{\theta}_*\right\|<\eta\right\}$ be an open neighborhood of $\theta^*$ and $-Q_t(\boldsymbol{\theta}):=-Q_{1 t}\left(\boldsymbol{\theta}^{(1)}\right)-Q_{2 t}\left(\boldsymbol{\theta}^{(2)}\right)$. Further, we would also need $\frac{1}{n-m} \sum_{t=m+1}^n E \left(-Q_t(\boldsymbol{\theta})\right)$ to have a unique minimum at $\bm{\theta}_0$ and that $\frac{1}{n-m} \sum_{t=m+1}^n E \left(-\sup_{\bm{\theta}\in\Theta}Q_t(\boldsymbol{\theta})\right)$ is finite. To simplify technical exposition, we assume A2 throughout this paper. 
    Assumption B2 ensures model identifiability. Combining subsequent assumption B3, Proposition \ref{Prop:preconditionsconsistency} will show assumptions B1 to B3 imply C1 to C2 for specific quasi-likelihood models. Assumptions C1 to C2 are needed to establish consistency as illustrated in Theorem \ref{Theorem:consistency}.
\end{remark}

\begin{theorem}\label{Theorem:consistency} Under Assumptions A2, B1, C1-C2, $\widehat{\boldsymbol\theta} \overset{p}{\to} \bm \theta_0$, as $n \to \infty$. 
\end{theorem}

The proof of Theorem \ref{Theorem:consistency} follows closely to the proof of Theorem 2.1 of \cite{zhu2011global} with the methods inspired by \cite{huber1967behavior}. Note that the above consistency result can be established for any pair of quasi-likelihood functions  $Q_{1\, t}(\bm \theta^{(1)})$ and $Q_{2\, t}(\bm \theta^{(2)})$ as long as they satisfy assumptions C1-C2. The key assumption for obtaining consistency is that the mean functions $\mu_{1\,t}$ and $\mu_{2\,t}$ are modeled correctly. 

Motivated by the real data application considered in this paper in Section \ref{sec:application}, we build $Q_{1\, t}(\bm \theta^{(1)})$ and $Q_{2\, t}(\bm \theta^{(2)})$ based on the variance function specifications $V_1(\mu_1)=\mu_1(1-\mu_1)$ and $V_2(\mu_2)=\mu_2$, respectively, which mimic the mean-variance relationship of the (for example) beta (parameterized in terms of mean and dispersion parameters) and double Poisson \citep{efron1986} distributions, which are used in practice for modeling counts and continuous bounded-valued data. Next, we provide verification of assumptions C1-C2 based on specific choices of $Q_{1\, t}(\bm \theta^{(1)})$ and $Q_{2\, t}(\bm \theta^{(2)})$. In other words, with the discrete series $Y_{2\,t}$ being modeled via double Poisson distribution and $Y_{1\,t}$ via Beta distribution, we have in the variance functions $V_1(\mu) = \mu(1-\mu)$ and $V_2(\mu) = \mu$. After omitting a few constant terms, we write 
\begin{eqnarray}\label{Eq:specuficQuasilikeli}
\begin{split}
&Q_1(y_{1\, t};\mu_{1\,t})\propto y_{1\, t}\log(\mu_{1\,t})+(1 - y_{1\, t})\log(1-\mu_{1\,t}),\\
&Q_2(y_{2\, t};\mu_{2\,t})\propto y_{2\, t}\log(\mu_{2\,t}) - \mu_{2\,t}.
\end{split}
\end{eqnarray}
Note that in order to ease notation usage, the dependency of $\mu_{1\,t}$ and $\mu_{2\,t}$ on $\bm \theta^{(1)}$ and $\bm \theta^{(2)}$ is suppressed. We also consider the link functions $g_1(x) = \log(\frac{x}{1-x})$ and $g_2(x) = \log(x)$. Next, we state an additional assumption required to establish the consistency of the estimators under the model specification given above. 

\begin{itemize}
    \item B3. Assume that
    \begin{gather*}
     E(Y_{2\,t}^2)< \infty , \;\;\max_{i=1,\ldots,p}E\left(|Y_{2\,t}Y_{2\,t-i}|\right)< \infty, \;\;  \max_{i=1,\ldots,p}E\left(Y_{2\,t-i}^2\prod_{j=1}^p(Y_{2\,t-j}^2+1)\right)< \infty, \\   
     \max_{i=1,\ldots,s}E\left(Y_{2\,t-i}^2\prod_{j=1}^s(Y_{2\,t-j}^2+1)\right)< \infty, \;\; \max_{i=1,\ldots,p} E\left(Y^2_{2\,t}Y_{2\,t-i}^2\prod_{j=1}^p(Y_{2\,t-j}^2+1)\right)< \infty. 
    \end{gather*}
\end{itemize}

Next, we state a proposition that shows that the consistency result in Theorem \ref{Theorem:consistency} holds for our case under Assumptions B1-B3.
\begin{proposition}\label{Prop:preconditionsconsistency}
   Suppose that Assumptions B1-B3 hold. Then, for the functions $Q_{1\, t}(\bm \theta^{(1)})$ and $Q_{2\, t}(\bm \theta^{(2)})$ described in  (\ref{Eq:specuficQuasilikeli}), Assumptions C1-C2 are satisfied. 
\end{proposition}

To establish the large sample distribution of the QMLE $\widehat{\bm \theta}$, we need a final additional technical assumption as follows. 
\begin{itemize}
    \item B4. The matrices $E  \left(\dfrac{\partial\nu_{1t}}{\partial\bm \theta^{(1)}}\dfrac{\partial\nu_{1t}^\top}{\partial\bm \theta^{(1)}}\right)$ and $E  \left(\dfrac{\partial\nu_{2t}}{\partial\boldsymbol\theta^{(2)}}\dfrac{\partial\nu_{2t}^\top}{\partial\boldsymbol\theta^{(2)}}\right)$ are both positive definite.
\end{itemize}


\begin{theorem}\label{Theorem:asymptotic}
Under assumptions B1-B4, we have 
\begin{eqnarray}\label{Eq:asymptoticRes}
\sqrt{n}(\widehat{\boldsymbol\theta}-\boldsymbol\theta_0)\stackrel{d}{\rightarrow}N({\bf0},\boldsymbol\Sigma),~\text{as}~ n\rightarrow\infty,
\end{eqnarray}
where $\boldsymbol\Sigma = \boldsymbol\Omega_2^{-1}\boldsymbol\Omega_1\boldsymbol\Omega_2^{-1}$, $$\boldsymbol\Omega_1\equiv \lim_{n\to\infty} E\left(n^{-1} \dfrac{\partial \widetilde{Q}(\boldsymbol\theta)}{\partial \bm \theta}\dfrac{\partial \widetilde{Q}(\boldsymbol\theta)}{\partial \bm \theta^{\top}}\bigg|_{\bm \theta = \bm \theta_0}\right); \boldsymbol\Omega_2\equiv \lim_{n\to\infty} E\left(n^{-1}\dfrac{\partial^2 \widetilde{Q}(\boldsymbol\theta)}{\partial \bm \theta \partial \bm \theta^{\prime}}\bigg|_{\bm \theta = \bm \theta_0}\right),$$ with the limits denoting limits in probability.

\end{theorem}


\begin{remark}\label{Remark:proofofnormality}
The proof of Theorem \ref{Theorem:asymptotic} is inspired by the proof technique utilized in Theorem 4.1.3 of \cite{amemiya1985advanced}. In practice, $\boldsymbol\Sigma$ can be consistently estimated by $\widehat{\boldsymbol\Sigma}={\bf S}_2^{-1}(\widehat{\boldsymbol\theta}){\bf S}_1(\widehat{\boldsymbol\theta}){\bf S}_2^{-1}(\widehat{\boldsymbol\theta})$, where ${\bf S}_1({\boldsymbol\theta})=n^{-1}\displaystyle\sum_{t=m+1}^n{\bf U}_t(\boldsymbol\theta){\bf U}_t(\boldsymbol\theta)^\top$ and ${\bf S}_2({\boldsymbol\theta})=n^{-1}\displaystyle\sum_{t=m+1}^n{\bf H}_t(\boldsymbol\theta)$, with ${\bf U}_t(\boldsymbol\theta)$ given by (\ref{eq:score}), and
\begin{eqnarray*}
{\bf H}_t(\boldsymbol\theta)&\equiv&E\left(-\partial{\bf U}_t(\boldsymbol\theta)/\partial\boldsymbol\theta\big|\mathcal F_{t-1}^{(1,2)}\right)\\
&=&
\begin{pmatrix}
\dfrac{1}{\phi_1V(\mu_{1t})(g_1'(\mu_{1t}))^2}\dfrac{\partial\nu_{1t}}{\partial\bm \theta^{(1)}}\dfrac{\partial\nu_{1t}^\top}{\partial\bm \theta^{(1)}} & {\bf0} \\
{\bf0} & \dfrac{1}{\phi_2V(\mu_{2t})(g_2'(\mu_{2t}))^2}\dfrac{\partial\nu_{2t}}{\partial\boldsymbol\theta^{(2)}}\dfrac{\partial\nu_{2t}^\top}{\partial\boldsymbol\theta^{(2)}} 
\end{pmatrix},
\end{eqnarray*}
with $\phi_1$ and $\phi_2$ replaced by consistent estimators if unknown. Therefore, the standard errors of $\widehat{\boldsymbol\theta}$ can be assessed via the matrix $n^{-1}\widehat{\boldsymbol\Sigma}$. To show that
    \begin{eqnarray}\label{Eq:convergenceS2}
    {\bf S}_2(\widehat {\bm \theta}) \overset{p}{\to} \boldsymbol\Omega_2,
    \end{eqnarray}
    the idea is to first argue that ${\bf S}_2({\boldsymbol\theta})=n^{-1}\displaystyle\sum_{t=m+1}^n{\bf H}_t(\boldsymbol\theta) \overset{p}{\to} \boldsymbol\Omega_2$ uniformly in a neighborhood of $\bm \theta_0$. Then, (\ref{Eq:convergenceS2}) holds for any sequence of estimators wherein $\widehat{\bm \theta} \overset{p}{\to} \bm \theta_0$. In other words, it is possible for us to consider ${\bf H}_t(\boldsymbol\theta)$ in a ``conditional expected version" of the hessian matrix (instead of just the hessian matrix) since we can estimate $\boldsymbol\Omega_2$ well by applying the tower property. As a result, the computation of the estimator for $\boldsymbol\Omega_2$ is simplified since the computation of all terms of the hessian matrix is cumbersome. 
\end{remark}

\subsection{Granger causality test}\label{sec:causality_test}
The interest here is in testing if the time series $\{Y_{1\,t}\}_{t\in\mathbb N}$ Granger causes $\{Y_{2\,t}\}_{t\in\mathbb N}$. This, in terms of competing hypotheses, can be formulated as $\texttt H_0: \gamma_l^{(2)} = 0\,\,\mbox{for all}\,\,l \in \{1,\ldots,k\}$, versus  $\texttt H_1: \gamma_l^{(2)}\neq0\,\,\mbox{for some}\,\,l \in \{1,\ldots,k\}$. The null hypothesis $\texttt H_0$ implies no Granger causality while the alternative hypothesis indicates Granger causality. 

We will use a likelihood ratio test based on quasi-likelihood to perform such a hypotheses testing. To do this, we define the vector of the bivariate time series and its associated conditional bivariate mean vector by $\bf{y}$ and $\boldsymbol\mu$, respectively, and the deviance function $D(\bf{y};\boldsymbol\mu)$ as two times the difference of the rescaled quasi-likelihood functions under the saturated (${\bf y}=\boldsymbol\mu$) and unsatured models:
\begin{eqnarray*}
D({\bf y};{\boldsymbol\mu})&=&2\left\{\sum_{t=m+1}^n\left[\phi_1Q_1(y_{1\,t};y_{1\,t})+\phi_2Q_2(y_{2\,t};y_{2\,t})\right]-\sum_{t=m+1}^n\left[\phi_1Q_1(y_{1\,t};\mu_{1\,t})+\phi_2Q_2(y_{2\,t};\mu_{2\,t})\right]\right\}\\
&=&-2\sum_{t=m+1}^n\left[\phi_1Q_1(y_{1\,t};\mu_{1\,t})+\phi_2Q_2(y_{2\,t};\mu_{2\,t})\right].
\end{eqnarray*}

Denote by $\widehat\mu$ and $\widehat\mu^0$ the quasi-likelihood estimates of the bivariate conditional mean vector under the unrestricted and restricted (null hypothesis) models. Then, the difference of the deviance under the restricted and unrestricted models is
\begin{eqnarray}\label{diff_dev}
D({\bf y};\widehat{\boldsymbol\mu}^0)-D({\bf y};\widehat{\boldsymbol\mu})=\phi_2\sum_{t=m+1}^n\left[Q_2(y_{2\,t};\widehat\mu_{2\,t})-Q_2(y_{2\,t};\widehat{\mu}_{2\,t}^0)\right].
\end{eqnarray}

Under the conditions of Theorem \ref{Theorem:asymptotic}, the rescaled (by $\phi_2$) quantity in (\ref{diff_dev}), denoted by quasi likelihood ratio (QLR) statistic, satisfies the following convergence in distribution:
\begin{eqnarray}\label{gc_test}
\mbox{QLR}&=&\sum_{t=m+1}^n\left[Q_2(y_{2\,t};\widehat\mu_{2\,t})-Q_2(y_{2\,t};\widehat{\mu}_{2\,t}^0)\right]\nonumber\\
&=&\dfrac{1}{\phi_2}
\sum_{t=m+1}^n\left[\int_{y_{2\,t}}^{\widehat\mu_{2\,t}}\dfrac{y_{2\,t}-\omega}{V_2(\omega)}d\omega-\int_{y_{2\,t}}^{\widehat{\mu}_{2\,t}^0}\dfrac{y_{2\,t}-\omega}{V_2(\omega)}d\omega\right]
\stackrel{d}{\longrightarrow}\chi^2_k,
\end{eqnarray}
as $n\rightarrow\infty$, where $\chi^2_k$ denotes a chi-squared distribution with $k$ degrees of freedom, with $\phi_2$ replaced by a consistent estimator if unknown. Note that we are not at the boundary of the parameter space under $H_0$ and therefore the asymptotic distribution of the QLR is indeed chi-squared. The convergence of the QLR statistic in (\ref{gc_test}) will be used in Section \ref{sec:application} to assess the Granger causality order in which the Covid-19 viral load influences mortality. In this case, the statistic assumes the closed-form 
$\mbox{QLR}=2\phi_2^{-1}\displaystyle\sum_{t=m+1}^n\left\{y_{2\,t}\log\left(\dfrac{\widehat{\mu}_{2\,t}^0}{\widehat{\mu}_{2\,t}}\dfrac{1-\widehat{\mu}_{2\,t}}{1-\widehat{\mu}_{2\,t}^0}\right)+\log\left(\dfrac{1-\widehat{\mu}_{2\,t}^0}{1-\widehat{\mu}_{2\,t}}\right)\right\}$, under the variance function choice $V_2(\mu_2)=\mu_2(1-\mu_2)$.

\section{Simulated Experiments}\label{sec:simulation}

This section assesses the estimation of the \texttt{MixTSQL} model parameters and their standard errors via Monte Carlo simulation. For the latter, we compare two approaches: one based on the asymptotic distribution of the QL estimators (Subsection \ref{subsec:asymptotics}), and an alternative pseudo-parametric bootstrap method, which involves: (i) fitting the model to the observed data; (ii) generating replicated trajectories using (i) and any distributional assumption that matches the two first moments QL specification; (iii) re-estimating parameters via quasi-likelihood for each replication; and (iv) computing standard errors as the standard deviation of the resulting estimates over $B$ replications. This pseudo-parametric bootstrap approach has been employed by \cite{maiaetal2021}, where a class of univariate semiparametric time series models was proposed based on quasi-likelihood and latent factor models.

A comparison between the theoretical and bootstrap approaches has been done previously in the context of count time series models in \cite{barreto2025time}. In the latter, the authors observed that standard errors are underestimated by the asymptotic method, and that the bootstrap is preferable under small samples. It is therefore essential to assess this behavior for the \texttt{MixTSQL} models, particularly given that our application involves a sample size of $n=112$ observations.

This study is carried out under four settings that vary in parameter values and distributional assumptions to generate time series trajectories. In Configurations 1 and 2 (C1, C2), the data is simulated from a bivariate beta-Poisson model (with conditional mean specifications (\ref{eq2_mu2}) and (\ref{eq1_mu1})) under the parameter values specified below and sample size $n=100$. 

\begin{itemize}
    \item \textcolor{blue}{Configuration 1:} (beta-Poisson) $\boldsymbol\beta^{(1)} = (1, 0.2)^\top$, $\boldsymbol\beta^{(2)} = (1, 0.2)^\top$, $\boldsymbol\gamma^{(1)} = -0.2$, $\boldsymbol\gamma^{(2)} = -0.2, \phi_1 = 0.2$, $\phi_2=1$. 

\item \textcolor{blue}{Configuration 2:} (beta-Poisson) $\boldsymbol\beta^{(1)} = (1.5, 0.2)^\top$, $\boldsymbol\beta^{(2)} = (1, 0.2)^\top$, $\boldsymbol\gamma^{(1)} = (-0.5, 0, 0, 0.3)^\top$, $\boldsymbol\gamma^{(2)} = (-0.2, 0, 0, 0.1)^\top$, $\phi_1 = 0.1$, $\phi_2=1$.

\end{itemize}

Analysis of the cross-correlation function (CCF) of the simulated data shows that these values of $\boldsymbol\gamma^{(1)}$ and $\boldsymbol\gamma^{(2)}$ render cross-correlation $\mbox{cor}(Y_{1,t}, Y_{2,t-1})$ and $\mbox{cor}(Y_{1,t-1}, Y_{2,t})$ of about $-0.30$ and $-0.17$, respectively. Fitting bootstrap SEs for C1 and C2 is carried out under correct model specification, i.e., both simulated data and the bootstrap replications are generated from the beta-Poisson model. This is changed in Configuration C3, where the data generation process and bootstrap distribution are no longer the same. 

In C3, the trajectories are simulated from a Bessel-Poisson model (Bessel distribution is an alternative to the beta model, and when parameterized in terms of the mean, matches the variance specification $\phi_2\mu_2(1-\mu_2)$; for instance, see \cite{bsetal2021}), and the pseudo-parametric bootstrap replications are gathered from a beta-Double Poisson specification. This analysis is important for evaluating the robustness and reliability of the pseudo-parametric bootstrap standard errors under model misspecification. Notably, the beta-Double Poisson specification used in the bootstrap replications offers the most flexible variance specification, as it includes nuisance parameters in both directions ($\phi_1$ and $\phi_2$). This added flexibility allows the model to accommodate different levels of dispersion, and we therefore expect it to perform well even when the assumed model does not match the true data-generating process. Parameter values employed in C3 are provided below.

\begin{itemize}
    \item  \textcolor{blue}{Configuration 3:} (bessel-Poisson) $\boldsymbol\beta^{(1)} = (-0.5, 0.2)^\top$, $\boldsymbol\beta^{(2)} = (1, 0.2)^\top$, $\boldsymbol\gamma^{(1)} = 0.25$, $\boldsymbol\gamma^{(2)} = 0.2$, $\phi_1 = 0.1$, $\phi_2 = 1$. 
\end{itemize}

\begin{figure}[ht!]
    \centering
    \includegraphics[width=0.9\linewidth]{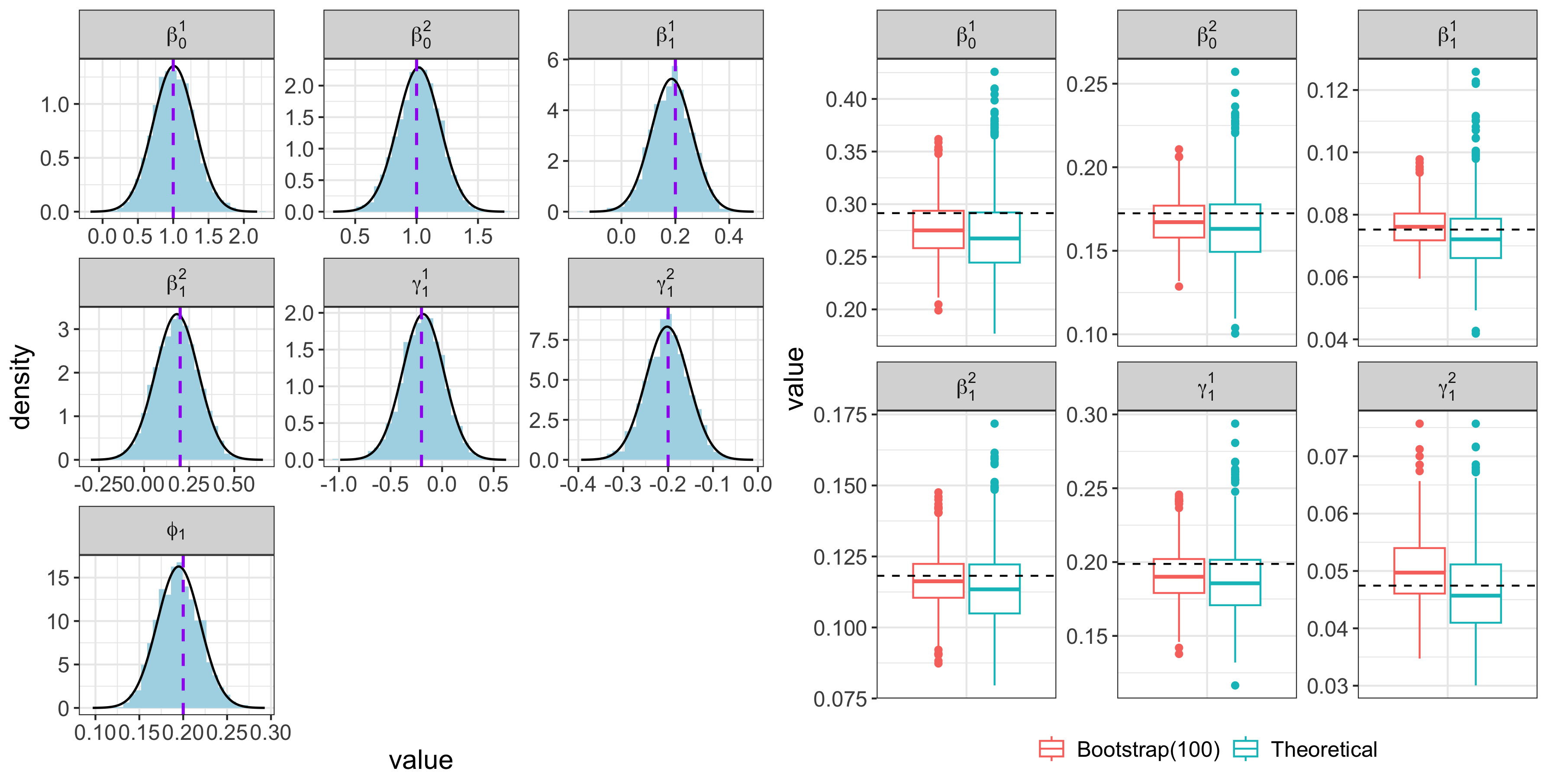}
    \caption{Estimation of model parameters and their standard errors under \textcolor{blue}{Configuration 1}. On the left, QMLE histograms with true values (vertical dashed lines) and fitted Gaussian curves. To the right, boxplots of bootstrap and theoretical standard errors. Horizontal dashed line indicates the Monte Carlo QMLEs standard deviation.} \label{fig:c1_results}
\end{figure}

Results from Configuration 1 are summarized in Figure \ref{fig:c1_results}. On the left, we display histograms of the quasi-maximum likelihood estimators (QMLEs) for each parameter. The vertical dashed lines indicate the true parameter values, and overlaid Gaussian curves illustrate the suitability of the normality assumption. The estimators are well-centered around the true values, and the normal approximation is appropriate even at such a small sample size. On the right, boxplots comparing standard errors estimated via bootstrap with 100 replications (denoted by Bootstrap(100)) and those derived from theoretical formulas are presented. A horizontal dashed line represents the Monte Carlo standard deviation of the QMLEs. Both bootstrap and theoretical standard errors are centered around this reference line, with a negligible difference between the two methods.

With Configuration 2, we explore the methods from a different perspective. The standard error estimates are now used to construct confidence intervals for $\gamma_1$ and $\gamma_2$. This enables evaluating their performance in model selection, specifically, in determining which cross-effects are statistically significant. For the bootstrap method, we examine two approaches to constructing confidence intervals: (i) using empirical bootstrap quantiles (2.5\% and 97.5\%) and (ii) combining the bootstrap SEs with normal quantiles. To assess the impact of replication size, we run the bootstrap procedure with $B=100$ and $B=500$. With the theoretical method, confidence intervals are constructed with Normal quantiles as in (ii). When fitting the model, we set a large number of lags, $k=s=10$, and construct confidence intervals (CIs). The idea is that this exercise mimics the fit of a full model to observed data, followed by the selection of relevant effects according to CIs.

\begin{figure}[ht!]
    \centering
    \includegraphics[width=0.8\linewidth]{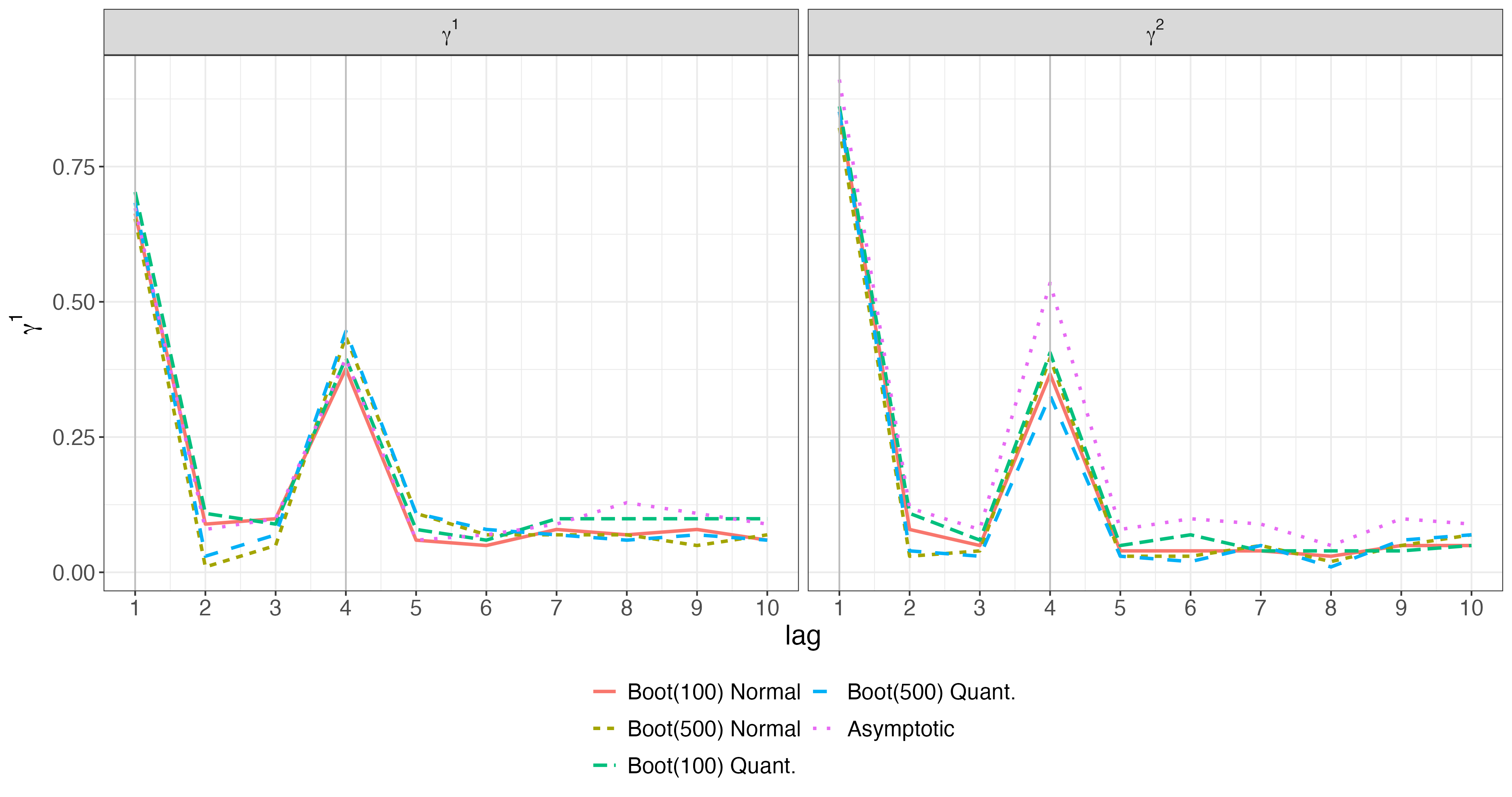}
    \caption{Detection of significant cross-effects under Configuration 2. Proportion of 1K simulations in which the confidence interval for $\gamma_1$ or $\gamma_2$ excluded zero, across different methods and lags. Results are shown for bootstrap intervals based on quantiles and normal approximations (with $B=100$ and $B=500$) and the theoretical method. True effects occur at lags 1 and 4, with a stronger signal at lag 1.} \label{fig:c2_intervals}
\end{figure}

The proportion of simulations in which the confidence interval for each effect excludes zero is reported in Figure \ref{fig:c2_intervals}. Ideally, significant cross-effects at true non-zero lags should be detected frequently. In this setup, the true effects occur at lags 1 and 4, where the cross-correlations are approximately -0.35 and 0.16, respectively. Results show that, as expected, detection rates are influenced by the signal strength. The stronger effect at lag 1 is correctly identified as statistically significant in approximately 80\% of simulations, while the weaker effect at lag 4 is detected around 50\% of the time. Across all settings, the results obtained using bootstrap-based confidence intervals are very similar to those from the theoretical method. Given this similarity, the theoretical approach may be preferable in practice due to its significantly lower computational cost.

\begin{figure}[ht!]
    \centering
    \includegraphics[width=0.9\linewidth]{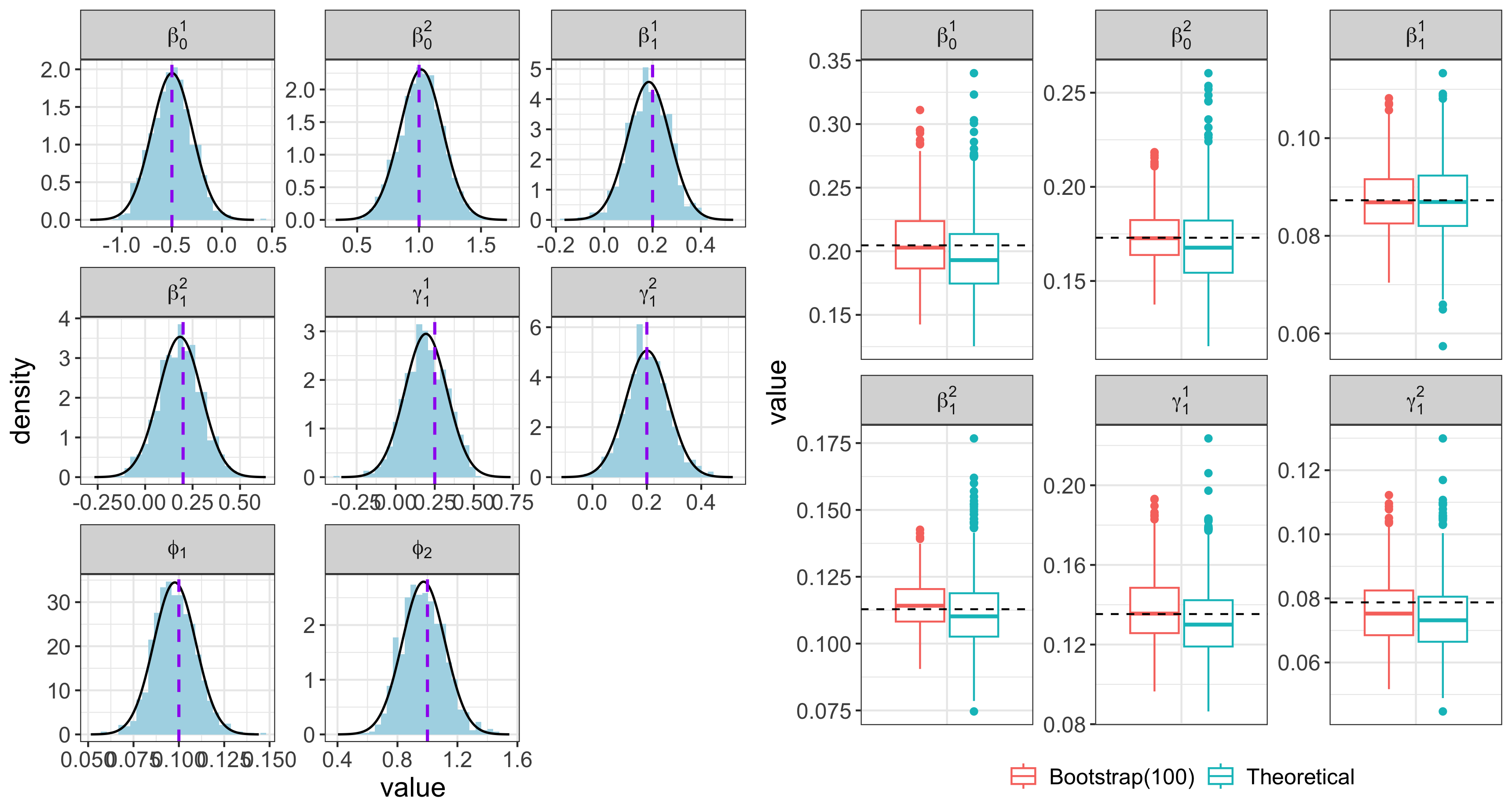}
    \caption{Estimation of model parameters and their standard errors under \textcolor{blue}{Configuration 3}. On the left, QMLE histograms with true values (vertical dashed lines) and fitted Gaussian curves. To the right, boxplots of bootstrap and theoretical standard errors. Horizontal dashed line indicates the Monte Carlo QMLEs standard deviation.} \label{fig:c3_results}
\end{figure}

The analysis presented in Figure \ref{fig:c1_results} is repeated for Configuration 3, where the data-generating process follows a Bessel-Poisson specification. In this case, the bootstrap procedure is based on beta-Double Poisson replicates. Figure \ref{fig:c3_results} shows that the mismatch between the bootstrap distributional assumption and data-generating mechanism does not affect the results. Standard errors are still well estimated and remain similar to the theoretical ones.

In conclusion, the estimation of standard errors via asymptotic theory performs well, even with a relatively small sample size of $n=100$. This has been confirmed under different data-generating processes and parameter settings. Comparisons with the bootstrap, using $B=100$ and $B=500$ replications, was carried out via pure SE estimation (C1) as well as confidence intervals (C2) and showed no apparent difference.

Given its low computational cost, the theoretical method will be preferable in most cases. A minor exception is if the standard errors of $\phi_1$ or $\phi_2$ are required, as these are not available from the theoretical method. However, their standard errors can be easily assessed from our pseudo-parametric bootstrap. For key tasks such as lag selection, the asymptotic SEs have been shown to be both reliable and computationally efficient, with no loss in accuracy compared to the bootstrap approach.


\section{Statistical Analysis of Covid-19 Viral Load and Mortality Dynamics}\label{sec:application}

We now provide a full statistical analysis of the Covid-19 dataset, by using the viral load to predict the number of future pandemic-related deaths, while also analyzing their causality relationship. To model the Covid-19 viral load ($Y_{1\, t}$) and its mortality - i.e., death counts ($Y_{2\, t}$), and explore their Granger causality, we consider the \texttt{MixTSQL} model 
\begin{equation}\label{eq:data_model}
\begin{aligned}
Y_{1\,t}|\mathcal F^{(1,2)}_{t-1} &\sim \mbox{QL}^{(1)}(\mu_{1\,t},\phi_1), \quad g_1(\mu_{1\,t}) = \beta_0^{(1)}+\sum_{i=1}^r \beta_i^{(1)}\widetilde Y_{1\, t-i}, \\
Y_{2\,t}|\mathcal F^{(1,2)}_{t-1} &\sim \mbox{QL}^{(2)}(\mu_{2\,t},\phi_2), \quad g_2(\mu_{2\,t}) = \beta_0^{(2)}+\sum_{i=1}^p \beta_i^{(2)}\widetilde Y_{2\, t-i}+\sum_{l=1}^k \gamma^{(2)}_l \widetilde Y_{1\, t-l},
\end{aligned}
\end{equation}
with variance functions $V_1(\mu_1)=\mu_1(1-\mu_1)$ and $V_2(\mu_2)=\mu_2$, $g_1(y) = T_1(y) = \log\left(\dfrac{y}{1-y}\right)$, $g_2(y) = \log y$ and $T_2(y) = \log(y+1)$, and $\phi_1\in(0,1)$ and $\phi_2>0$ being unknown dispersion parameters, to be estimated using Equations (\ref{eq:phi_est}). The above specification models causality in the direction from viral load to death counts, not the other way around. The sample size consists of $n=112$ weekly observations.

\begin{figure}[ht!]
    \centering
    \includegraphics[width=0.95\linewidth]{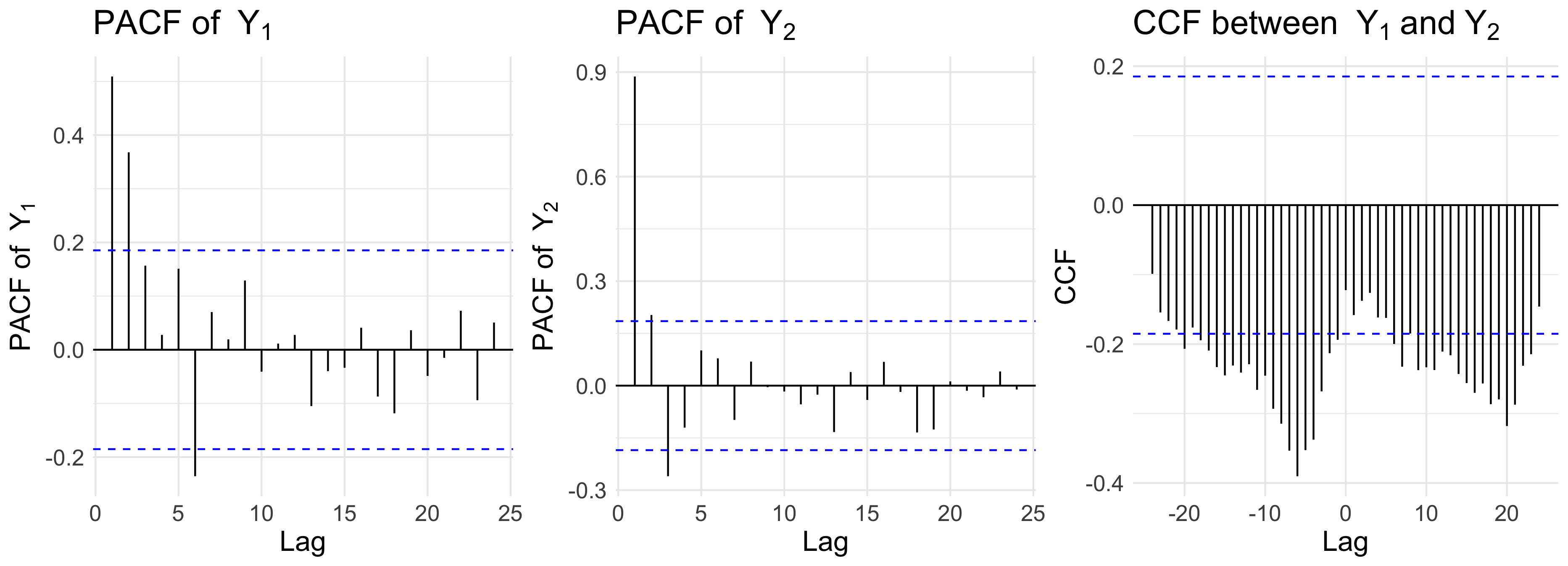}
\caption{Sample PACF and CCF plots for the weekly time series \(Y_{1\,t}\) (viral load) and \(Y_{2\,t}\) (death counts).}\label{fig:acfpacf}
\end{figure}

Preliminarily, the marginal autocorrelation function (ACF) and partial ACF (PACF) of the two series are inspected, as well as their cross-correlation function (CCF). We use these plots to guide the initial setup of the autoregressive orders \((p, r)\) and the causality order \((k)\). The PACF and CCF are reported in Figure \ref{fig:acfpacf}. We consider the maximum lag as 24, which covers the relationship between two series spanning half a year. Along with the ACF, the PACF suggests an autoregressive model of order 6 for \(Y_{1\,t}\) and order 3 for \(Y_{2\,t}\), as indicated by prominent peaks at these lags in the PACF and a gradual decay in the ACF, characteristic of AR processes.

Model~(\ref{eq:data_model}) is fitted with autoregressive orders \(r = 6\) for \(Y_{1\,t}\) and \(p = 5\) for \(Y_{2\,t}\), including all lags in between. The maximum cross-correlation order is set to \(k = 10\). Our goal is to use this as an initial full model and let confidence intervals of the estimated parameters guide the model specification. To this end, the theoretical standard errors derived in Subsection \ref{subsec:asymptotics} are computed. By using asymptotic 95\% confidence intervals, the following effects are deemed statistically significant: for \(Y_{1\,t}\), autoregressive lags 1,2,5 and 6; for \(Y_{2,t}\), autoregressive lag 2 and cross-lagged influence from \(Y_{1\,t}\) at lag 6. All coefficients remain statistically significant in this reduced model, for which parameter estimates and 95\% confidence intervals are reported in Table \ref{tab:app}.

\begin{table}[ht]
\centering
\begin{tabular}{|ccc|}
\hline
Parameter & Estimate & 95\% CI \\
\hline
$\beta_0^{(1)}$  & 0.058 & ($-$0.074, 0.189) \\
$\beta^{(1)}_1$     & 0.261 & (0.084, 0.438) \\
$\beta^{(1)}_2$      & 0.262 & (0.112, 0.413) \\
$\beta^{(1)}_5$     & 0.226 & (0.096, 0.355) \\
$\beta^{(1)}_6$     & $-$0.186 & ($-$0.313, $-$0.058) \\
$\beta^{(2)}_0$   & 1.161 & (0.444, 1.878) \\
$\beta^{(2)}_2$     & 0.801 & (0.686, 0.915) \\
$\gamma^{(2)}_6$    & 0.122 & (0.045, 0.198) \\
\hline
\end{tabular}
\caption{Parameter estimates with 95\% confidence intervals.}\label{tab:app}
\end{table}

The finding of the lagged effect (after six weeks) of the viral load on death counts confirms an expected result at the aggregate level. The rationale is that it should take a few weeks from the point that a patient is contaminated by the virus to the corresponding hospitalization and, depending on the corresponding severity level, the death of the patient. This is also supported by previous studies, in which at the individual level, viral load peaks during the presymptomatic period and subsequently declines, resulting in progressively higher Ct values as the interval between infection acquisition and specimen collection lengthens \citep{lin2022incorporating, jones2021estimating}.

\begin{figure}[ht!]
     \centering
     \includegraphics[width=0.9\linewidth]{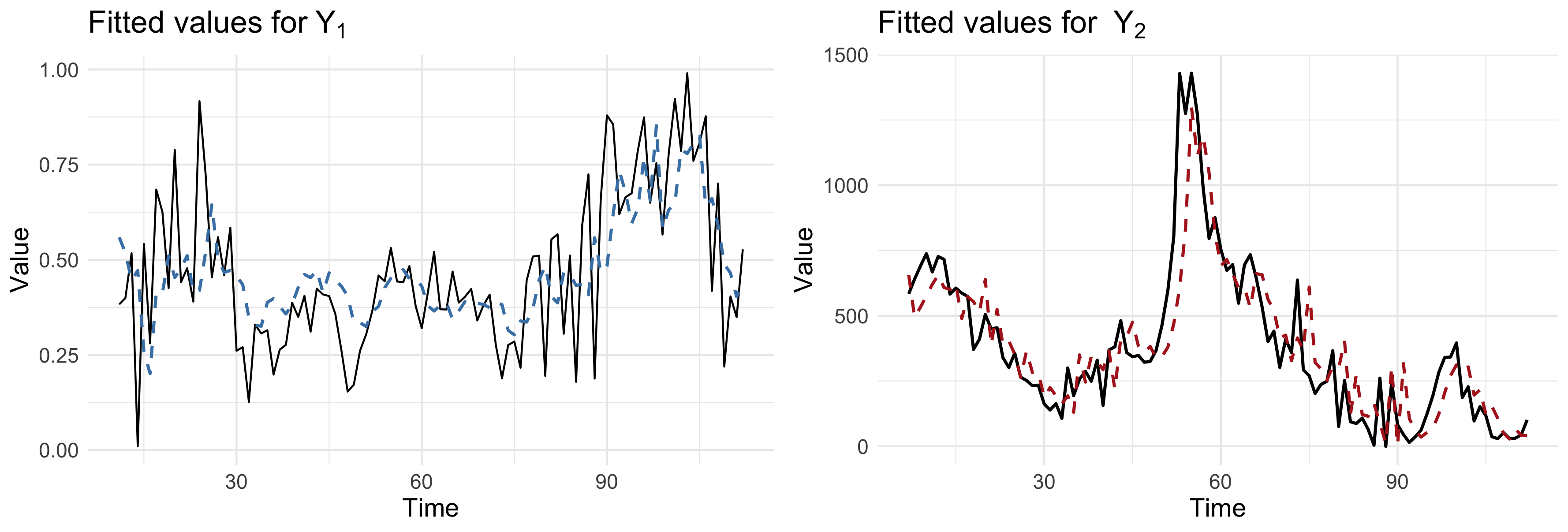}
     \caption{Observed series \(Y_{1\,t}\) and \(Y_{2\,t}\) (solid lines) and their corresponding fitted values \(\widehat{\mu}_{1\, t}\) and \(\widehat{\mu}_{2\, t}\) (dashed lines).}\label{fig:fitted_vals}
 \end{figure}

A visual inspection of the fit of the model to the data (in-sample prediction) is provided in Figure \ref{fig:fitted_vals}. The estimated conditional means of \(Y_{1\, t}\) and \(Y_{2\, t}\), i.e. \(\widehat{\mu}_{1\, t}\) and \(\widehat{\mu}_{2\, t}\), are plotted alongside the observed trajectories. Notably, the fitted values (dashed lines) show a good match to the observed trajectories (solid lines).

Importantly, Figure \ref{fig:fitted_vals} also reflects that both fitted values appropriately capture the three most relevant epidemiological waves of the Covid-19 pandemic in Brazil. More specifically, the first wave occurred from May to September 2020. This was primarily driven by early ancestral SARS-CoV-2 lineages - B.1 and its derivatives, leading to widespread mortality at a time when vaccination was unavailable and health systems experienced considerable strain. The second wave occurred between late 2020 and mid-2021, being marked by the emergence and rapid dissemination of the Gamma variant P.1 \citep{zeiser2022first, giovanetti2022replacement}. This lineage displayed increased transmissibility and immune escape potential, culminating in the deadliest phase of the pandemic in Brazil. The third wave was defined by the global spread of Omicron and its subvariants, while progressively expanding vaccination coverage and naturally acquired immunity among the population. This wave occurred from December 2021 to March 2022, dominated by early Omicron subvariants, such as the BA.1. This produced unprecedented surges in case incidence, although with a comparatively attenuated mortality, underscoring the protective effects of prior immunization \citep{arantes2023comparative, berra2024covid}.

\begin{figure}[ht!]
    \centering
    \includegraphics[width=0.9\linewidth]{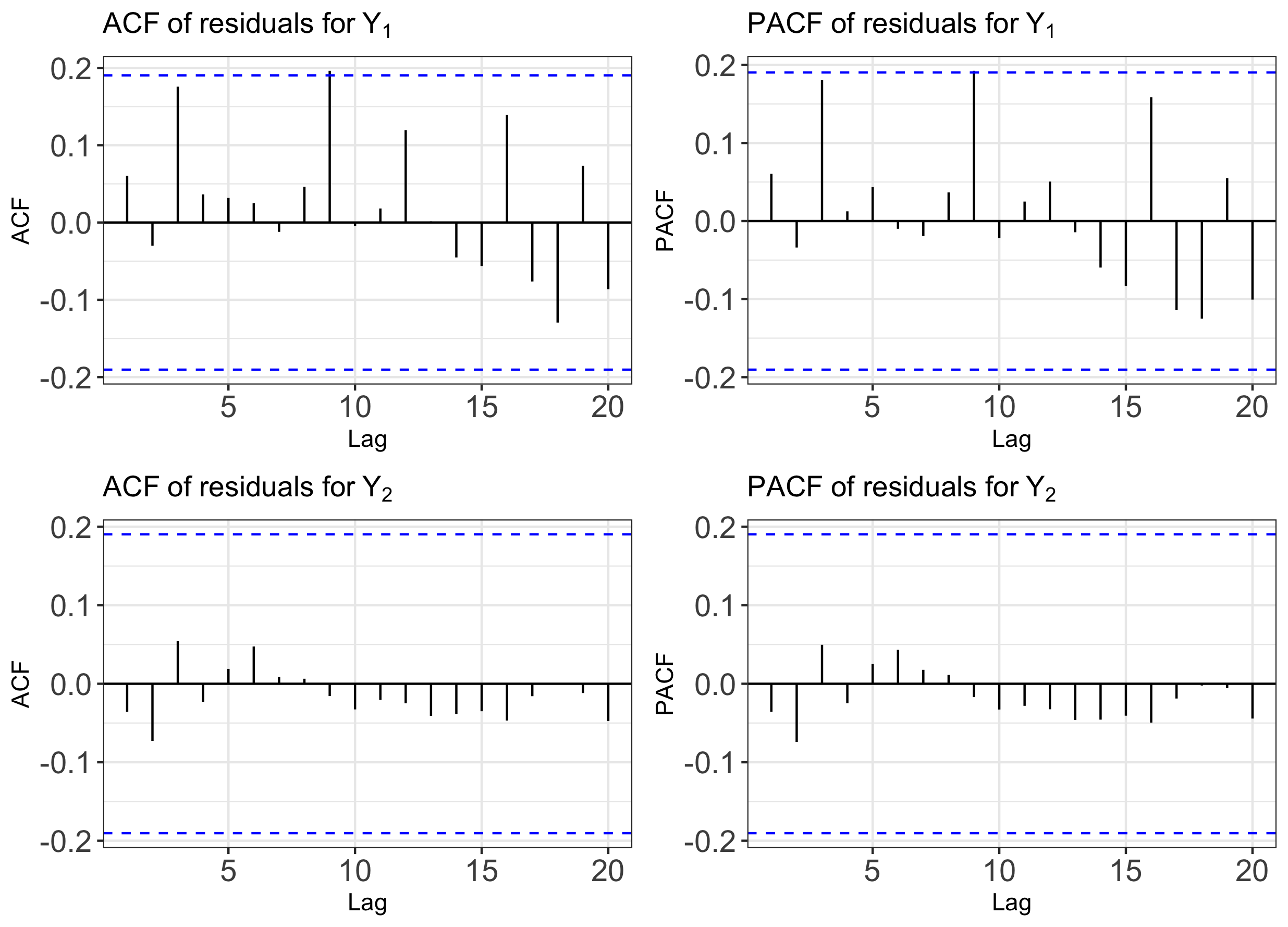}
   \caption{Sample autocorrelation (ACF) and partial autocorrelation (PACF) plots of the residuals from the fitted models for \(Y_{1,t}\) and \(Y_{2\,t}\).}
   \label{fig:acfpacfofresidual}
\end{figure}

Subsequently, we assess the fitted model’s goodness of fit and forecasting performance. Adequacy of the stipulated dynamics and variance function are examined through residuals check and probability integral transform (PIT) plots. Since no specific distributional assumptions are imposed in our methodology, we assume auxiliary distributions that match the first two moments in order to construct a PIT plot. This allows us to assess whether the assumed mean–variance relationship is adequate. Prediction performance is evaluated through a one-step-ahead out-of-sample forecasting exercise. First, residual autocorrelation is inspected using the ACF and PACF plots reported in Figure~\ref{fig:acfpacfofresidual}. No significant autocorrelation is identified, indicating that the models adequately capture the dynamics of the two series. Given adequate fitted dynamics, inspection of the PIT plot provides an outlook of the distributional assumptions of count models. This is a well-cemented diagnostic tool for integer-valued data used to evaluate if the observed dispersion behaviour is being well captured. We consider a PIT analysis of the death count trajectory, which is computed as follows.


\begin{figure}[ht!]
    \centering
    \includegraphics[width=0.9\linewidth]{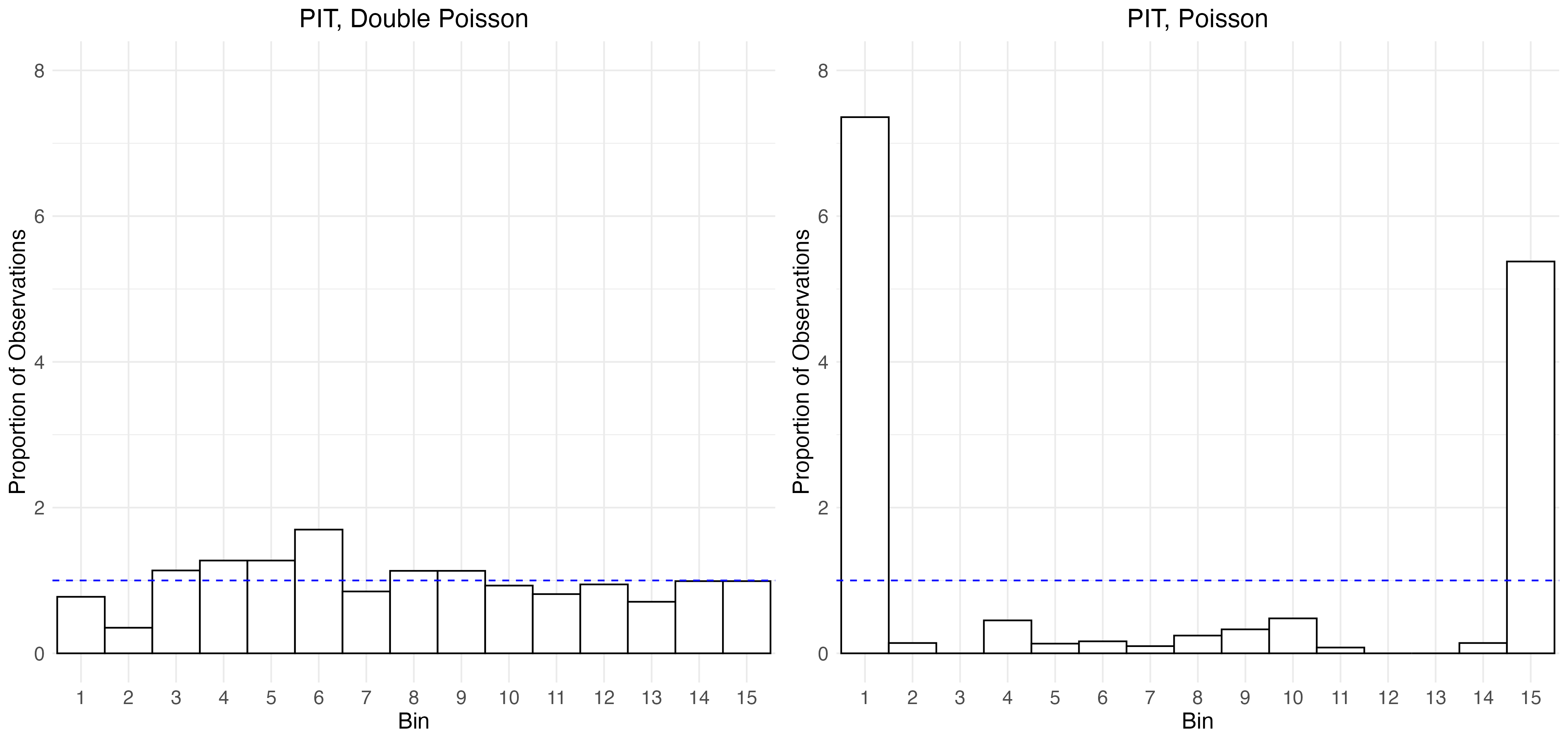}
\caption{Probability integral transform (PIT) histograms for $Y_2$. The left panel shows the PIT based on the Double Poisson model, while the right panel shows the PIT based on the Poisson model. The dashed blue line at one indicates the expected uniform value for a well-calibrated predictive distribution.} \label{fig:PIT}
\end{figure}

By definition, a PIT diagnostic evaluates $\mbox{PIT}_t = \widehat{F}_t(y_t)$, where $y_t$ is the observed data at time $t$ and $\widehat{F}(\cdot)$ is the fitted cumulative distribution function (CDF). In fully parametric models, assuming conditional Poisson or negative binomial distributions, for example, $\widehat{F}(\cdot)$ corresponds to the CDF of $y_t$ at the estimated parameters. The PIT plot is a histogram where the [0,1] interval is divided into bins, and the proportion of $\mbox{PIT}_t$ falling into each bin is computed. The histogram should be approximately uniform when well-calibrated. Deviations from uniformity indicate potential misspecification, for example, U-shaped histograms indicate that the fitted distribution is underdispersed because the observations fall more often in its tails. Figure \ref{fig:PIT} displays the PIT plots under the double Poisson and Poisson specifications. The key distinction between these two models, in terms of their mean–variance relationship, lies in the inclusion of an additional dispersion parameter in the double Poisson distribution. From the figure, it is evident that the double Poisson model provides an adequate fit, whereas the standard Poisson model does not. Consequently, the variance function $V(\mu_2)=\mu_2$ with dispersion parameter $\phi_2\neq 1$ is appropriate for modeling the mortality data.

For the Granger causality, we test the hypotheses $H_0: \gamma_6 =0$ (no Granger causality) against $H_1: \gamma_6 \neq0$ ($Y_{1\, t}$ Granger causes $Y_{2\, t}$). As per Subsection \ref{sec:causality_test}, the test is carried out using the QLR statistic. We obtain the test statistic of 20.21, which gives strong evidence (p-value $< 10^{-5}$) to reject $H_0$. In other words, the test supports that the viral load $Y_{1\, t}$ Granger causes the mortality $Y_{2\, t}$.

Beyond the investigation of goodness of fit, we are also interested in the prediction performance. Specifically, the model is first estimated using observations from a training sample comprising the initial 50 weeks of data. Let $T$ denote the number of observations in the training set, initially $T=50$. Model (\ref{eq:data_model}) is fitted, and we compute the point forecast $\widehat{\mu}_{2,T+1}$ for the number of deaths at $T+1$. As a proxy for 95\% prediction intervals, we employ the 2.5\% and 97.5\% quantiles based on a double-Poisson model (which matches our mean-variance specification). This procedure is carried out recursively - at each iteration, the training sample is extended by one observation, the model is re-estimated, and the corresponding prediction is recorded. For benchmarking purposes, we perform the same exercise using a Gaussian time series model with square-root–transformed response. The latter keeps the same autoregressive and causal lag structure. This recursive evaluation yields 62 one-step-ahead forecasts from each model, which we compared against the observed values.

\begin{figure}[ht!]
    \centering
    \includegraphics[width=0.85\linewidth]{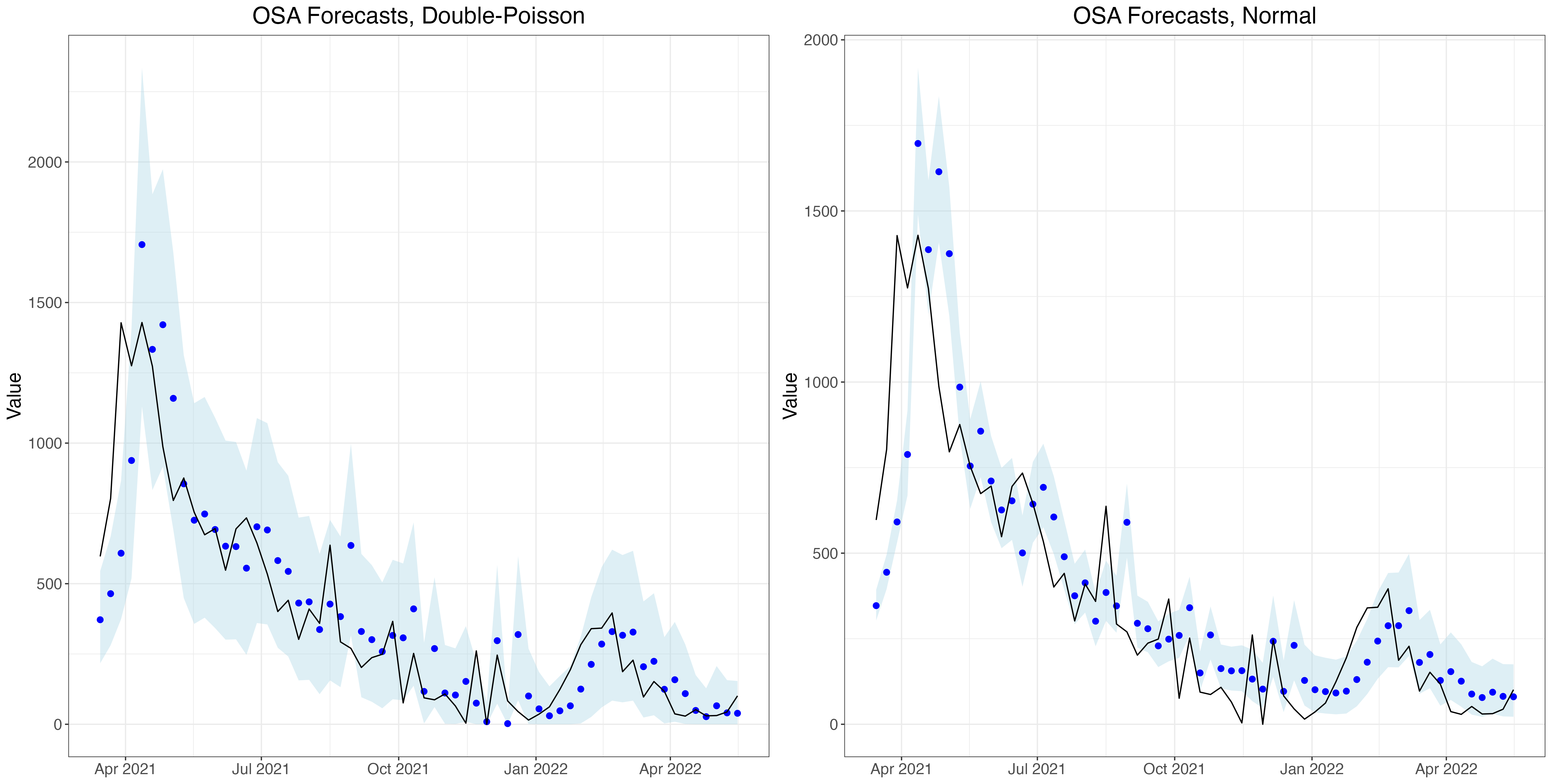}
    \caption{One-step-ahead (OSA) forecasts from the double-Poisson model (left) and the Gaussian model (right). 
    The black line represents the observed series, blue dots correspond to OSA predictions, and the shaded area denotes the 95\% prediction interval.} \label{fig:OSA_forecasts}
\end{figure}

\begin{figure}[ht!]
    \centering
    \includegraphics[width=0.6\linewidth]{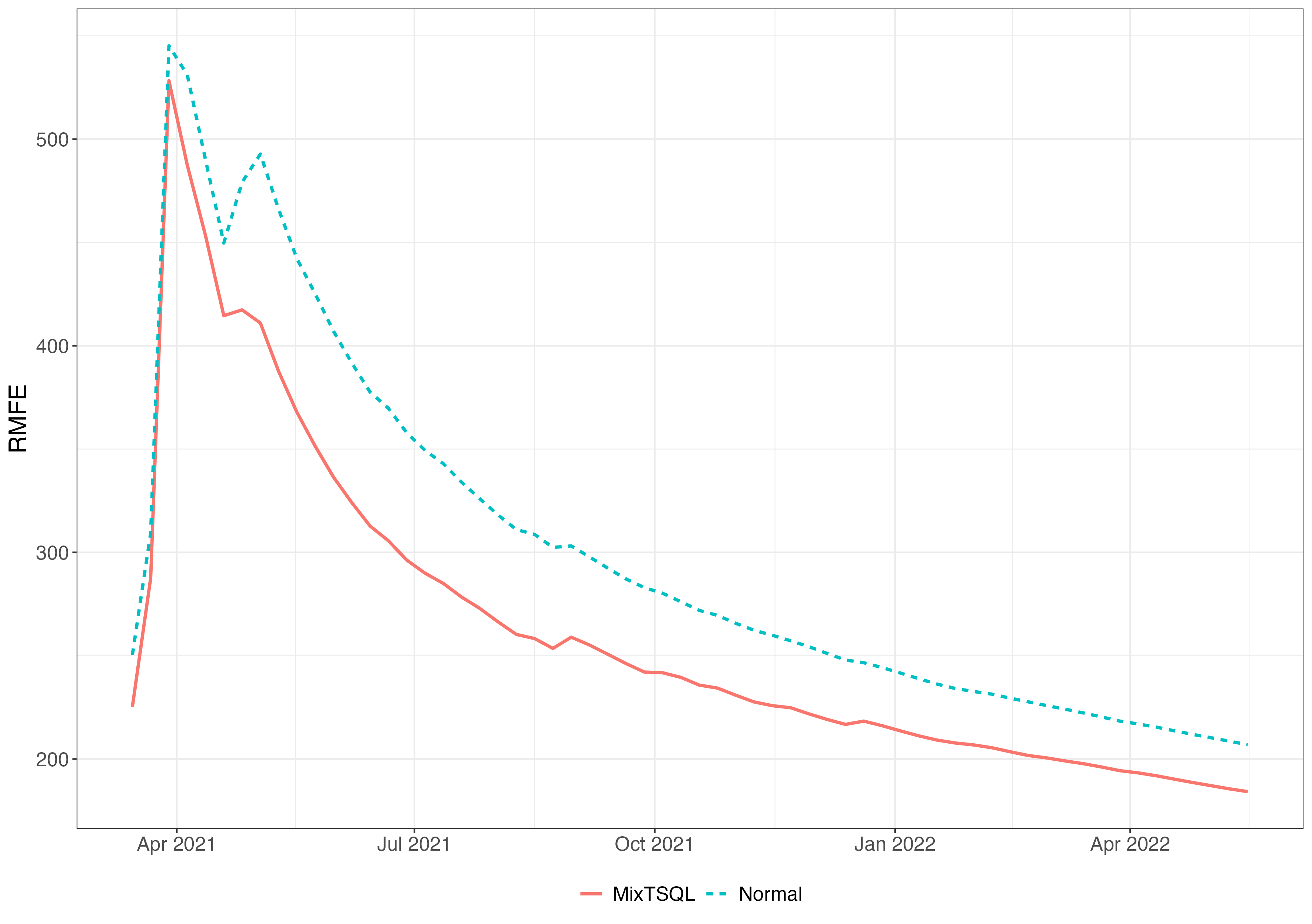}
    \caption{Root Mean Forecasting Error (RMFE) of out-of-sample forecasting exercise based on the \texttt{MixTSQL} and Gaussian models.}  \label{fig:RMFE}
\end{figure}

Results are illustrated in Figure \ref{fig:OSA_forecasts} and Figure \ref{fig:RMFE}. In the first, the observed data are shown against point predictions and their confidence intervals. Observations are represented with a black solid line, while point predictions are given as dots with their 95\% bands shown as shaded areas. Notably, the Normal model produces narrower confidence bands but these often do not comprise the true counts. In Figure \ref{fig:RMFE}, the prediction error from each model is quantified using the root mean forecasting error (RMFE). The RMFE is defined as the square root of the mean squared forecasting error, that is, 
\[
\text{RMFE}_H = \left( \frac{1}{H} \sum_{h=1}^{H} \left( y_{T+h} - \widehat{y}_{T+h} \right)^{2} \right)^{1/2},\quad H=1,\ldots,n-T,
\]  
where $y_{T+h}$ denotes the realized outcome at time $h$, $\widehat{y}_{T+h}$ the corresponding point forecast, and $H$ the total number of predictions. Thus, the RMFE provides a cumulative measure of the average prediction accuracy across the evaluation window.

The \texttt{MixTSQL} specification renders a lower RMFE relative to the Gaussian linear model. This highlights that not only does the proposed model maintain the original scale of the data, but also produces more accurate short-horizon mortality forecasts. In this application, variance-stabilizing transformations such as the square root are essential for employing in the Gaussian model. We also note that results from the log transformation are similar.

\section{Concluding remarks}
\label{sec:conclusions}

This paper was motivated by the joint modeling of the Covid-19 viral load and its associated mortality. To handle the nature of this mixed bivariate time series, we introduced a novel quasi-likelihood modeling framework, termed \texttt{MixTSQL}, which enables the analysis of mixed-valued time series comprising binary, count, positive, continuous, and bounded-valued variables. Designed specifically to address the complexities of real-world epidemiological data, our \texttt{MixTSQL} models only require the specification of the conditional mean and variance, avoiding restrictive distributional assumptions and enabling flexibility.

We also developed a Granger causality test, which is tailored to mixed-valued data, filling a significant methodological gap in the literature. From a theoretical perspective, we established the asymptotic properties of the quasi-maximum likelihood estimators, ensuring theoretical robustness and inferential reliability. Extensive simulation studies further confirmed the effectiveness of \texttt{MixTSQL} models in terms of estimation accuracy, with both theoretical and bootstrap-based inference methods performing well in finite samples. The consistency between asymptotic results and empirical findings strengthens confidence in the method’s applicability to a broad range of problems.

\texttt{MixTSQL} models were employed to study the dynamic relationship between Covid-19 viral load (measured via Ct values) and mortality in Brazil. Our analysis reveals a statistically significant Granger-causal effect of viral load on future deaths, highlighting the potential of Ct values/viral load as leading indicators in pandemic surveillance. This application not only validates the practical utility of \texttt{MixTSQL} models but also underscores its relevance for real-time public health decision-making, especially in the context of epidemic control.

Promising directions for future research include incorporating additional public health covariates, such as vaccination coverage and mobility restrictions, into the modeling framework, and reconstructing viral load proxies using widely available indicators like the number of cases and deaths. Methodologically, enhancing uncertainty quantification around the estimated conditional mean and developing multi-step ahead forecasting techniques within the \texttt{MixTSQL} framework would further broaden its scope and impact across applied domains.

{\footnotesize

\section*{Data Availability Statement}
Data are available from the authors upon request.

\section*{Conflict of interest}
The authors do not have competing interest to be declared.
}


\spacingset{1}
{\scriptsize
\bibliographystyle{Chicago}
\bibliography{Bibliography}
}

\end{document}